\documentclass[usenatbib]{mn2e}
\usepackage{aas_macros,psfig,amsmath}
\newcommand{\neff}{n_{\mathrm{eff}}}
\newcommand{\alphaeff}{\alpha_{\mathrm{eff}}}
\newcommand{\kappaeff}{\kappa_{\mathrm{eff}}}

\begin{document}
\topmargin-1cm
\title{Precision Cosmology from the Lyman-$\alpha$ Forest: Power Spectrum and Bispectrum}

\author[R. Mandelbaum, P. McDonald, U. Seljak, R. Cen]{
R.~Mandelbaum$^1$, %\thanks{rmandelb@feynman.princeton.edu}, 
P.~McDonald$^1$, %\thanks{pmcdonal@feynman.princeton.edu}, 
U.~Seljak$^1$, %\thanks{uros@feynman.princeton.edu},
R.~Cen$^2$\\%\thanks{e-mail here} \\
$^1$ Department of Physics, Princeton University, Princeton, NJ 08544,
 USA \\$^2$ Department of Astrophysical Sciences, Princeton University, Princeton, NJ 08544,
 USA}

\pubyear{2003}

\maketitle

\begin{abstract}
We investigate the promise of the Ly-$\alpha$ forest for high 
precision cosmology in the era of the Sloan Digital Sky Survey 
using low order N-point statistics.
We show that with the existing data one can determine the 
amplitude, slope and curvature of the slope of the matter 
power spectrum with a few percent precision.  
Higher order 
statistics such as the bispectrum provide independent information 
that can confirm and improve upon the statistical precision from the 
power spectrum alone. The achievable precision is comparable to 
that from
the cosmic microwave background with upcoming satellites, 
and complements it by measuring 
the power spectrum amplitude and shape at smaller scales. Since the data 
cover the redshift range $2<z<4$, one can also extract the evolution of
the growth factor and Hubble parameter over this range, and provide
useful constraints on the presence of dark energy at $z>2$.

\end{abstract}

%\keywords{large-scale structure of the universe; cosmological parameters}
%\pacs{PACS here}

\section{Introduction}\label{S:intro}

The study of the Ly-$\alpha$ forest has been revolutionized in recent years
by high resolution measurements using the Keck HIRES spectrograph \citep{1994SPIE.2198..362V}
and by the development of
theoretical understanding using hydrodynamical simulations 
\citep{1994ApJ...437L...9C,1995ApJ...453L..57Z,1996ApJ...457L..51H,1998MNRAS.301..478T}
and analytical models \citep{1998MNRAS.296...44G}.
The picture that has emerged from these studies
is one in which the neutral gas responsible for the absorption is in a relatively
low density, smooth environment, which implies a simple connection between the gas
and the underlying dark matter. 
The neutral fraction of the gas is determined
by the interplay between the recombination rate (which depends on the
temperature of the gas) and ionization caused by
ultraviolet photons. Photoionization heating
and expansion cooling cause the gas density and temperature to be tightly
related, except where mild shocks heat up the gas. This
leads to a tight relation between the absorption
and the gas density. 
Finally, the gas density is
closely related to the dark matter density on large scales, while on small
scales the effects of thermal broadening and Jeans smoothing have to
be included.
In the simplest picture described here all of the physics ingredients are
known and can be modeled.

Within the model described above,
the relation between the observed flux and the underlying dark 
matter is completely defined, enabling one to study the dark matter 
correlations with the help of the Lyman-$\alpha$ forest.
This was proposed first by \cite{1998ApJ...495...44C} and has been subsequently 
investigated and applied to the real data by several groups 
\citep{1999ApJ...520....1C,2002ApJ...581...20C,2000ApJ...543....1M,2001ApJ...552...15H}. 
This work established the 
simple flux power spectrum as the statistic of choice,
although 
other statistics such as the flux probability 
distribution have also been investigated \citep{2000ApJ...543....1M}.  

This paper addresses in more detail
the question of how much information about cosmological parameters
can be extracted from the analysis of the Lyman-$\alpha$ forest 
and what are the best statistics to use. We 
focus on N-point statistics (where N=2, 3, 4), since these provide 
the simplest parametrization of the long-range correlations. 
%%%This part of the paragraph added in response to referee's comments.
We use a Fisher matrix analysis to 
assess the uncertainties  
in cosmological parameters for a given uncertainty in Lya forest  
measurements. To calculate the Fisher matrix, we
must calculate the derivatives of the observables (flux power spectrum  
and bispectrum) with respect to cosmological parameters,  
centered on a reasonable fiducial model. 
Using this approach, %%%this is the end of what I added
we show that 
higher order statistics add to and independently
confirm the information gained from the flux power spectrum, and we
address practical issues of such a study.
For the data sample, we assume a sample with noise and 
resolution characteristics of the Sloan Digital Sky Survey, 
which currently contains a few thousand QSO spectra with 
measured Lyman-$\alpha$ forest ($z>2.2$).  

Statistical correlations in the Lyman-$\alpha$ forest are sensitive 
to several parameters of cosmological interest. Broadly, 
they are sensitive 
to the linear power spectrum of matter fluctuations on scales 
around 1 comoving $h^{-1}$Mpc. 
In this paper,
this power spectrum $P_{L}(k, a)$ is expressed as 
\begin{equation}\label{E:Pdef}
P_{L}(k, a) = A D^2(a) k^{n+(\alpha/2) \ln{(k/k_p)}} T^2(k).
\end{equation}
The parameters that are being studied are the power spectrum amplitude 
($\kappa \equiv \ln{\sqrt{A}}$), the primordial slope $n$, and
primordial curvature $\alpha$. 
We also investigate the sensitivity 
to the linear growth factor $D(a)$, where $a=(1+z)^{-1}$ is the expansion 
factor.  

Several additional parameters affecting absorption in the Lyman-$\alpha$ forest
must be included, such as those of the gas temperature-density relation, 
which for our redshift range is assumed to have the form
\begin{equation}
T(\rho) = T_{1.4} \left( \frac{\rho}{1.4\bar{\rho}} \right)^{\gamma-1}.
\end{equation}
$T_{1.4}$ is used because \citet{2001ApJ...562...52M} determined the
temperature at this density most precisely.
$T_{1.4}$ and $\gamma$ are two of the parameters studied in this paper.
In addition, the mean transmitted flux $\bar{F}(z)$ is also assumed 
to be a free parameter (which can be related to the UV 
background and baryon density though the equation of ionizing 
equilibrium).
Thus six parameters were varied at each redshift.
Our approach is to use hydrodynamical and N-body simulations, and 
vary all the parameters at each redshift. We are interested 
in the sensitivity to cosmological parameters, so we marginalize over
all the other parameters when presenting the results.

The outline of the paper is the following. 
In section \ref{S:sims} we describe our HPM simulations and the hydrodynamic 
simulations that are used for comparison, introduce the statistics used in 
the course of the analysis, and discuss our choice of simulations, filters 
and instrumental effects
for this study.
Section~\ref{S:fisher} describes the procedure of calculating the
Fisher matrix, and the
determination of errorbars for the parameters of interest.
In section \ref{S:quintessence}, we address the possibility of using the
Lyman-$\alpha$ forest to determine quintessence density $\Omega_q$ or
equation of state $w_q$, and more general deviations from the Einstein-de Sitter
growth factor $D(a)=a$.
After discussing the implications of the results in section \ref{S:discussion}, we conclude with a discussion of future work in this area.

\section{Simulations}\label{S:sims}
%%%Moved this paragraph to here, from section 3, due to referee's comments.
To assess the sensitivity of the Ly-$\alpha$ forest to cosmological parameters
we performed a Fisher matrix analysis (e.g. \citet{1997ApJ...480...22T}).
We conducted a  redshift-dependent analysis 
using 9 redshift bins, centered at $z=2.2, 2.4, \ldots, 3.8$ and using 
a realistic redshift distribution of quasars as found in SDSS data.  
Since we only had simulations with values of $a$ spaced by 0.04, we chose the closest $a$ available.%, and used a value of $\bar{F}$ from the fit in Eq. \ref{E:Fz}.

We use Hydro-PM (HPM) simulations \citep{1998MNRAS.296...44G,2001astro.ph..8064M,2002ApJ...580...42M,2002ApJ...570..457C} rather than more accurate hydrodynamic simulations due to the very large number of simulations needed.  
The boxes used for the main analysis had length 80 $h^{-1}$Mpc, with $N=512^3$ particles, and spectra were averaged over
all three axes.   
Pixelization, smoothing, and noise of the data were imitated in the simulations, with $S/N = 5$ and Gaussian smoothing $\sigma$ of length 0.625 comoving $h^{-1}$Mpc (same size as the pixels), as will be described in more detail in section \ref{S:noise}.
  $64^2$ lines of sight into each face of the box were used to reduce computation time after finding that using $128^2$ did not yield more information. 

%%%Made this a new paragraph; before, it was part of the previous one.
The fiducial model used for the full Fisher analysis is $\gamma-1=0.3$, $T_{1.4}=17000$ K, $n=0.75$, $\alpha=0.0$, $\Omega_{m,0}h=0.26$, $\sigma_8 = 0.75$
and $\bar{F}(z)$ from  
\begin{equation}\label{E:Fz}
\bar{F}(z) = e^{-e^{-0.95+3.77\ln{[(1+z)/4.0]}}}, 
\end{equation}
a fit to the data in \citet{2000ApJ...543....1M}.  
Note that $n$ and $\alpha$ here are the primordial slope and
curvature.
The values measured are $\neff$ and $\alphaeff$, which include
a contribution from the transfer function.  
The value chosen for $\gamma-1$ is the center of the range of
theoretical values, and two-sided derivatives were done in steps of
$\pm 0.03$ (in other words,  using
simulations with $\gamma-1 = 0.0$ and 0.6).  Likewise, the value chosen
for $T_{1.4}$ is observationally justifiable, and derivatives were
taken in steps of $\pm 5000$ K.   Derivatives in $n$ about 0.75 and
$\alpha$ about 0.0 (a value consistent with observation as well) were
both done in steps of $\pm 0.05$.  Derivatives in $\bar{F}(z)$ were in
steps of $\pm 0.02$.  
%%%Added this from Fisher matrix section because it is also about the derivatives.
Instead of varying the actual power spectrum amplitude, which requires a large number of different simulations, we varied $a$.  
In relating derivatives such as $\mathrm{d}P/\mathrm{d}a$ to the
desired derivatives $\mathrm{d}P/\mathrm{d}\kappa$, we used the
approximation (shown in \citet{2001astro.ph..8064M} to be quite
accurate) that $z$-evolution of the power spectrum can be treated as a
rescaling of $\kappa$ and $T_{1.4}$.  The change in growth factor can
be treated primarily as a change in $\kappa$, along with a small change in $T_{1.4}$ (fixed in velocity
coordinates, it consequently differs in
simulation coordinates).
%%%Added this sentence for clarity
Two-sided derivatives with respect to $a$ were calculated using steps of $\pm 0.04$.
Then, by subtracting off this temperature term from the
$a$-derivatives, we obtain the portion of the $a$-derivative that
mimics a change in power spectrum amplitude evolution.  Then we use
$\mathrm{d}\kappa = \mathrm{d}a/a$ to form the amplitude derivative
(this form assumes an Einstein-de Sitter universe with $D(a)=a$).

While the bulk of our simulations are HPM, we use 
the hydrodynamic simulations for comparison whenever possible. These
 are described in \citet{2001ApJ...559L...5C,2002ApJ...580...42M,2002ApJ...570..457C}; and for more detail, \citet{2002astro.ph..3524C}.   
They have box size 25 $h^{-1}$Mpc, with $768^3$ Eulerian cells for baryons 
and $384^3$ dark matter particles, and the cosmology is $\Lambda$CDM, with $\Omega_{\Lambda} = 0.7$, $\Omega_{m}=0.3$, $\Omega_b=0.035$, $\sigma_8 = 0.9$, $h=0.67$, and $n=1$.  
Median temperature at the mean density is around 15,000 K, slightly 
decreasing with redshift for $2<z<4$,  
consistent with observational constraints in \cite{2000ApJ...543....1M}.
The three outputs used have redshifts 1.9, 2.45, and 3.0.  
The HPM simulations used for the comparison with hydrodynamic simulations had 
a smaller box size than those used for the full analysis.  
The parameters were box size 25~$h^{-1}$Mpc, $384^3$ particles, the same
cosmology as the hydrodynamic simulations, and $\gamma-1$=0.6 (the temperature-density relation in the hydrodynamic simulations is not precisely a power law, but the slope is near 0.6).  
The two had nearly identical initial conditions, minimizing the sampling 
variance.
The missing mass that was concentrated into
stars in the hydrodynamic simulations was neglected in the HPM
simulations.
In all cases, $\bar{F}=0.7$ was used for the comparison, with 2-sided derivatives using increments of $\pm 0.02$.  
The simulations were binned so that the hydrodynamic and HPM
simulations had the same size pixels.  

We now introduce the statistics used for analysis, and then use them
to address several questions about these simulations: accuracy in
comparison with hydrodynamic simulations, sufficiency of the box size
and resolution, the magnitude of random fluctuations due to finite box
size, and the effects
of chunking on analysis.  We discuss all but the last of these issues only to determine the
sufficiency of our simulations for a Fisher matrix study, rather than
requiring the higher precision necessary for their use in a data
analysis.  The final issue, chunking, is discussed since it may be 
done in a data analysis, so it is necessary to show that it does 
not affect the results.

\subsection{Statistics}\label{S:statistics}
In general, the statistics were computed using the flux residual
$\delta_F = F/\bar{F}-1$ rather than $F$ itself.  
Besides the flux power spectrum $P$ with normalization
\begin{equation}
\int_{-\infty}^{+\infty} \frac{dk}{2\pi} P(k) = \langle \delta_F^2 \rangle,
\end{equation}
several higher order statistics were
studied.  Rather than investigate the full bispectrum and trispectrum 
information, we decided to use a more restricted form of these 
general statistics. In this sense our results will be conservative. 
The higher order statistics are created using the following
procedure: $\delta_F(\lambda)$ is Fourier transformed to find
$\tilde{\delta}_{F}(k)$.  
Next, it is band-pass filtered to create the field
\begin{equation}
\tilde{\delta}_H(k) = \tilde{\delta}_F(k)W(k)
\end{equation}
where $W(k)$ is some filtering function. Two types of filters used in this analysis are a square window,
\begin{equation}\label{E:SquareW}
W_s(k) = 
\begin{cases}
0, &\text{if $k<k_1$ or $k>k_2$} \\
1, &\text{if $k_1 \le k \le k_2$}
\end{cases}
\end{equation}
and a Gaussian window characterized by some $\bar{k}$ and $\sigma_k$.

Next, the filtered field $\tilde{\delta}_H(k)$ is inverse transformed to create 
the real-space counterpart, $\delta_H(\lambda)$.  The field $H(\lambda)$ is created by squaring:
\begin{equation}
H(\lambda) = (\delta_H(\lambda))^2.
\end{equation}
We define 
$T$ as the power spectrum of $H$.
Another statistic, $B$, is the cross-spectrum between
$\delta_{F}$ and $H$.  $B$ and $T$ are
related to the three- and four-point functions, respectively.
However, because $T$ is not the reduced four-point function, there is
a significant covariance between it and $P$.

In addition, the cross-correlation coefficient
\begin{equation}
R = \frac{B}{\sqrt{PT}}
\end{equation}
was studied by \cite{2001ApJ...551...48Z} due to its simple physical 
interpretation. 
As shown there, it is expected that
$R$ computed from the flux residual should be significantly negative
for low $k$, since perturbations in high density (low flux) regions 
tend to grow faster.  As shown in this paper,
one utility of these higher order statistics is
that they 
break degeneracies between the power spectrum amplitude and $\bar{F}$ as
measured by $P$ alone.  Furthermore, as shown in
\citet{2001ApJ...551...48Z}, they may also be useful in discriminating
between gravitational processes (which give negative $B$ and $R$) and
continuum fluctuations (which give positive $B$ and $R$) or other
extraneous effects such as metal lines, star-formation induced outflows
of gas, inhomogeneity in the UV background, and the limitations of simulations.

\begin{figure}
\centerline{\psfig{file=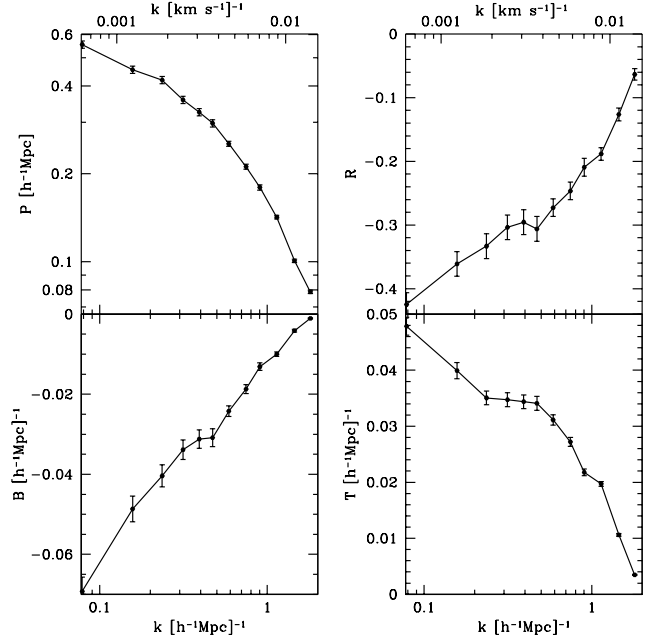,width=3.5in}}
\caption{The values of the statistics for the fiducial model noted in
the text.  Errorbars
are those expected for the amount of data in this redshift bin.  The filter is B1 (see Table~\ref{T:filters}), a square filter from 0.2-1.1 $h$Mpc$^{-1}$.}
\label{F:fiducial}
\end{figure}

Fig. \ref{F:fiducial} shows $P$, $B$, $T$, and $R$ for the central
point in parameter space around which variation occurs for $z=3.0$
($\gamma-1=0.3$, $T_{1.4}=17000$ K, $a = 0.24$, $n=0.75$,
$\alpha=0.0$, $\bar{F}=0.68$).  Note that we vary $a$ rather than the
power spectrum amplitude itself (which involves generating many more
simulations). By Eq. \ref{E:Pdef}, we have $\delta \kappa = \delta
a/a$ for an Einstein-de Sitter cosmology ($D(a)=a$).  Errorbars shown are
computed using the variances of the statistics for a single simulation
and the length of spectrum for this redshift in the SDSS data
available to us.  

\begin{figure}
\centerline{\psfig{file=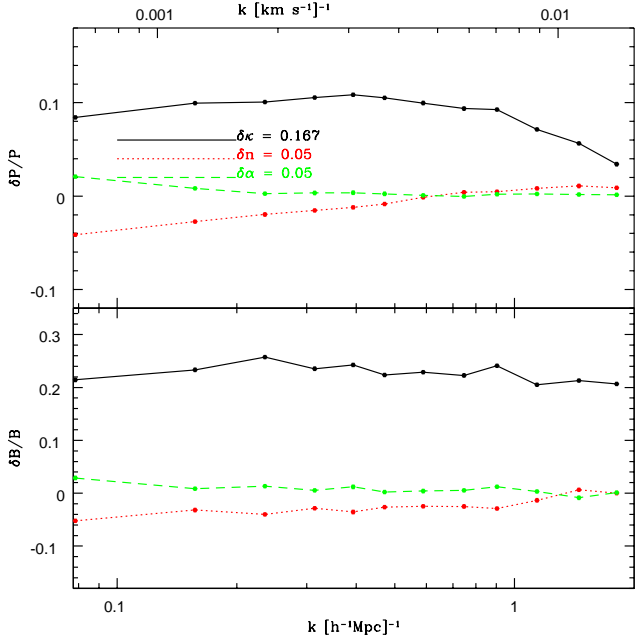,width=3.5in}}
\caption{Plot of the relative variation of the statistics with $n$,
$\alpha$, and $\kappa$ for one square
filter.}
\label{F:powparams}
\end{figure}

\begin{figure}
\centerline{\psfig{file=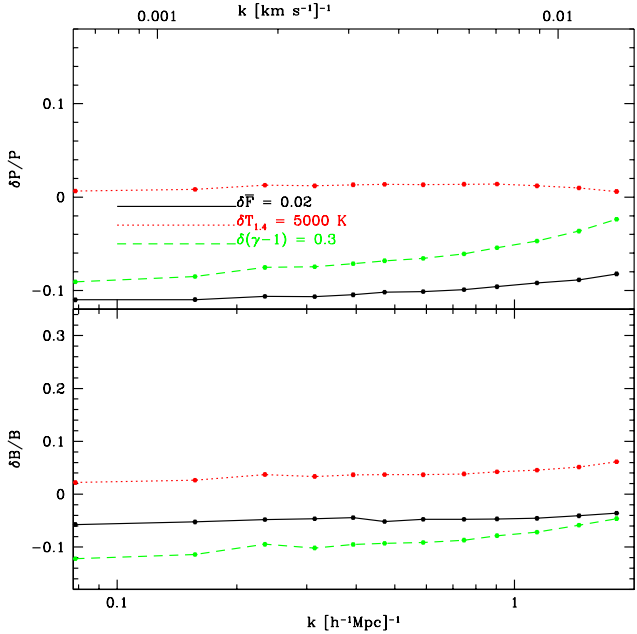,width=3.5in}}
\caption{Plot of the relative variation of the statistics with
$\bar{F}$, $\gamma-1$, and $T_{1.4}$.}
\label{F:otherparams}
\end{figure}

Our study showed that $T$ and $R$ do not add much information, since
$T$ is so closely related to $P$, and $R$ is just a combination of the
other statistics.  
Consequently, in the rest of this paper, only $P$ and $B$ will be
used.  
Figs. \ref{F:powparams} and \ref{F:otherparams} show the relative variation of
these statistics with all 6 parameters.  Note that in this section,
all plots of the values of statistics or their derivatives were
created using the same filter, a square filter from 0.2-1.1
$h$Mpc$^{-1}$ (unless noted otherwise), and derivative plots have the same scale for easy comparison.

%%%This paragraph added to address referee's comments.
We briefly explain the effects of changing the various parameters on
$P$ and $B$ that are illustrated in Figs.~\ref{F:powparams}
and~\ref{F:otherparams}.  Raising the overall power spectrum amplitude
causes an increase in $P$ and $B$ on all scales shown here, because the increase in
power in the matter power spectrum carries directly over to 
the flux power spectrum.  Increasing $n$ lowers $P$ and $B$ on large
scales and raises it on small scales, directly reflecting its effects on
the matter power spectrum (for even smaller scales than those shown
here, increasing the power spectrum amplitude and $n$ actually causes
a decrease in $P$ because non-linear
peculiar velocities suppress power, as in
\citet{2001astro.ph..8064M}).  The statistics are nearly insensitive to
$\alpha$ because it does not have much effect on these scales.  
 $P$ and $B$ show only a very
slight increase with $T_{1.4}$.  The decrease of these statistics with
an increase in $\gamma-1$, consistent with \citet{2001astro.ph..8064M}, can be explained by considering the
optical depth $\tau \propto \rho_b^{2-0.7(\gamma-1)}$, so that the
exponent on $\rho_b$ is decreasing, and hence the fluctuations in optical depth
for a given density fluctuation also decrease.  The decrease in power
with an increase in the mean flux is consistent with the results in
\citet{2001astro.ph..8064M}, and can be understood by the proportionality
between the optical depth and $\rho_b^{2-0.7(\gamma-1)}$.
Because the increased mean flux is equivalent to a lower optical
depth, we can see that this should have the same effect as 
lowering the density fluctuations, and consequently the flux
power spectrum (hence the degeneracy between power spectrum amplitude and mean flux).

\subsection{Comparison of HPM and Hydro simulations}

\begin{figure}
\centerline{\psfig{file=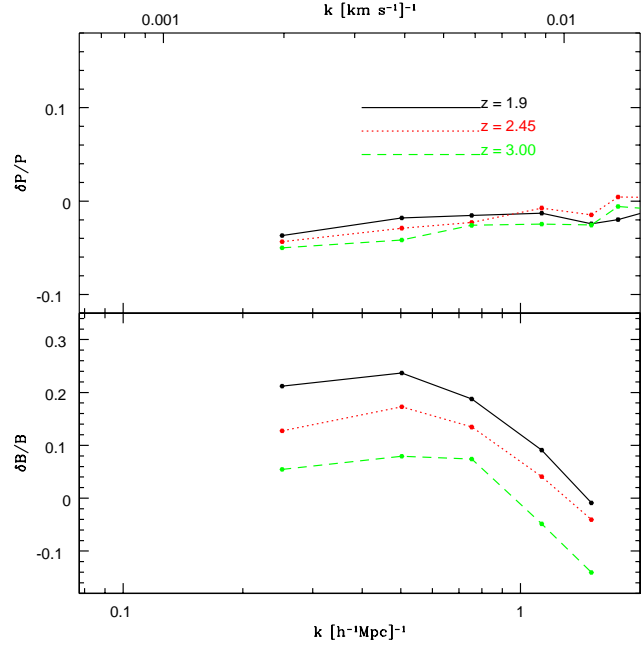,width=3.5in}}   
\caption{Relative difference in $P$ and $B$ for hydrodynamic vs. HPM
simulations, for three $z$ values.}
\label{F:hydrohpm}   
\end{figure}

\begin{figure}  
\centerline{\psfig{file=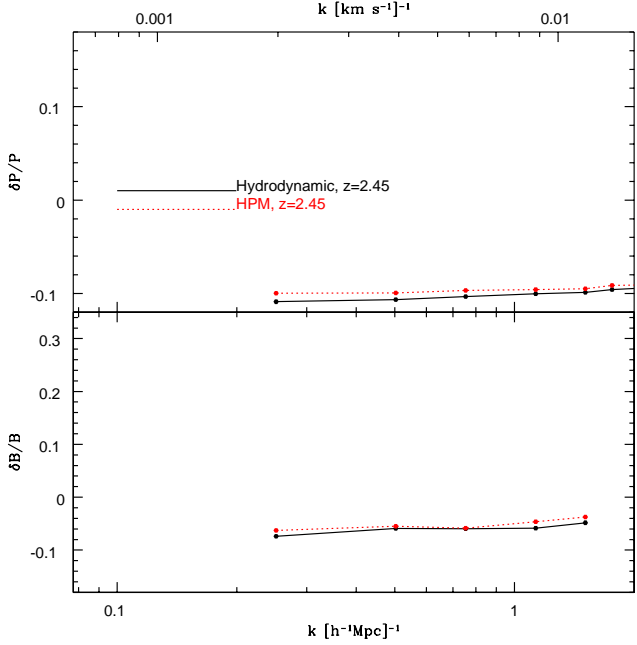,width=3.5in}}   
\caption{Relative difference in two-sided $\bar{F}$-derivatives 
hydrodynamic vs. HPM simulations, at $z=2.45$,
with $\delta \bar{F} = 0.02$.}
\label{F:hydrohpmdF}   
\end{figure}

\begin{figure}    
\centerline{\psfig{file=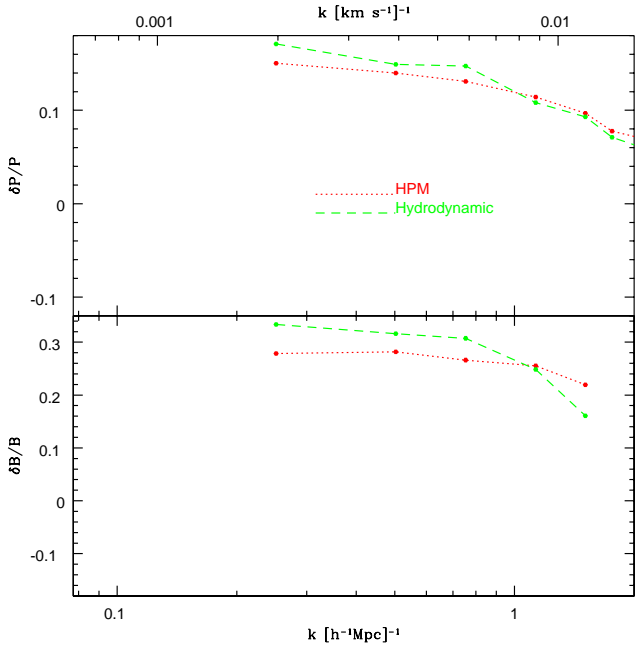,width=3.5in}}    
\caption{Comparison of amplitude derivatives for hydrodynamic vs. HPM simulations, with $\bar{F}=0.7$
and $\delta \kappa =0.167$.} 
\label{F:hydrohpmda}    
\end{figure}

Fig.~\ref{F:hydrohpm} shows the relative difference between HPM and
hydrodynamic simulations at three redshifts, Fig.~\ref{F:hydrohpmdF} shows
their flux derivatives at $z=2.45$, and
Fig.~\ref{F:hydrohpmda} shows their amplitude derivatives at
$z=2.45$.  

For the derivatives, the relative change (e.g., $\delta P/P$) was used for comparison, because differences in the magnitude of the derivatives could be simply artifacts of the differences in the magnitudes of the statistics themselves.  
Since this kind of change will cancel out of a Fisher matrix
calculation (see equation \ref{E:fishdef}), the relative change
is plotted to emphasize changes in the $k$-dependent shape of the
derivatives rather than their amplitude.  

As illustrated by Figs. \ref{F:hydrohpm}-\ref{F:hydrohpmda}, the HPM 
simulations yield statistics and derivatives fairly similar to those of
the hydrodynamic simulations.  The worst discrepancies in
Fig.~\ref{F:hydrohpm} occur for $z=1.9$, already below the range of
this study. 
A Fisher matrix analysis using just amplitude and $\bar{F}$ was
completed for these simulations, and found that the error bounds for
both parameters were 6\% lower when calculated with the HPM 
than with the hydro simulations, a tolerable difference for our purposes. 

\subsection{Resolution and box size}
\begin{figure}    
\centerline{\psfig{file=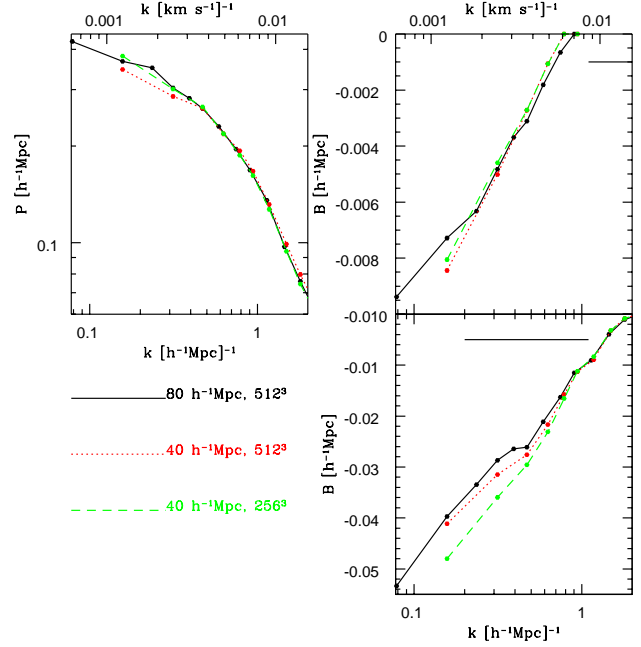,width=3.5in}}    
\caption{Comparison of statistics for different box size and
resolution, with filters 0.2-1.1 $h^{-1}$Mpc (bottom) and 1.1-2.0
$h^{-1}$Mpc (top) as shown on plot.} 
\label{F:res}    
\end{figure}

\begin{figure}    
\centerline{\psfig{file=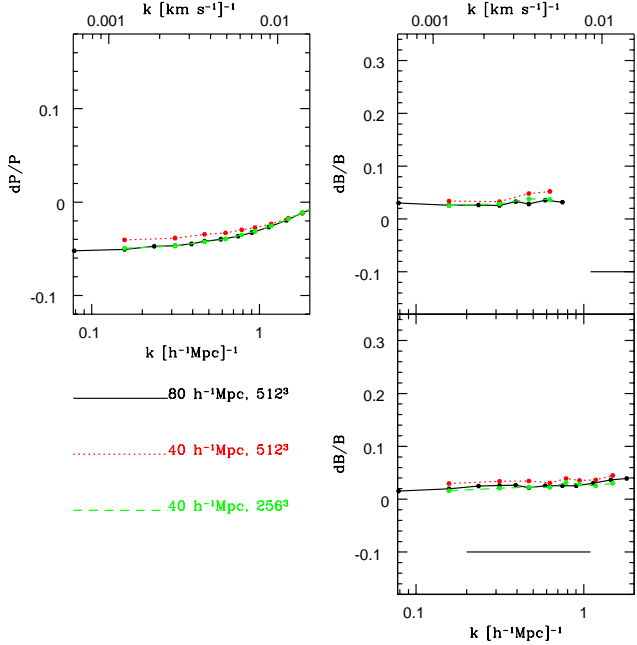,width=3.5in}}    
\caption{Comparison of $\bar{F}$-derivatives ($\delta \bar{F}=0.02$) for different box size and resolution, as noted on plot. The choice of filters 
for $B$ is the same as in previous plot.} 
\label{F:resf}    
\end{figure}

\begin{figure}    
\centerline{\psfig{file=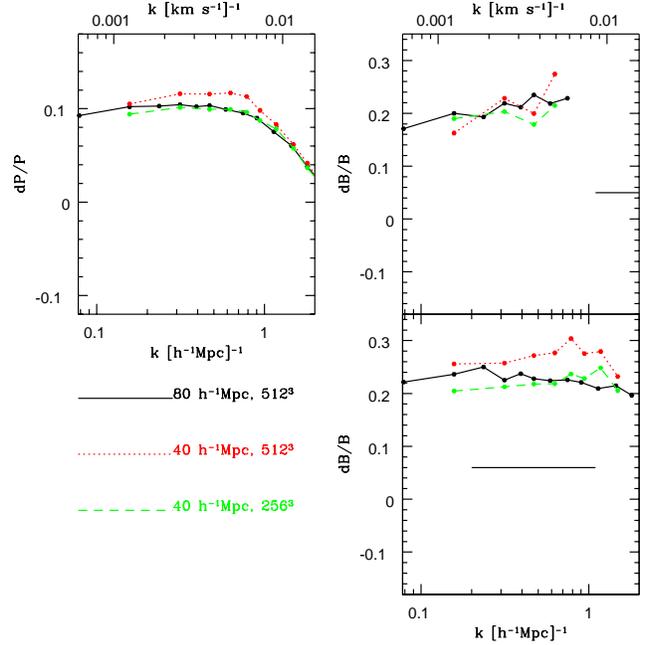,width=3.5in}}    
\caption{Comparison of $\kappa$-derivatives ($\delta \kappa=0.167$) for different box size and resolution, as noted on plot. The choice of filters 
for $B$ is the same as in previous plot.} 
\label{F:resa}    
\end{figure}

Another question about the HPM simulations is the box size and resolution necessary for a realistic study of these statistics.  
To check whether the 80 $h^{-1}$Mpc box with $512^3$ particles is
sufficient, we computed the statistics and their derivatives for a 40
$h^{-1}$Mpc box with $512^3$ particles 
(higher resolution), and a 40 $h^{-1}$Mpc box with $256^3$ particles
(same resolution, different box size).  
The values of the statistics and their derivatives with respect to
$\bar{F}$ and $\kappa$ with $\gamma-1=0.3$, $T_{1.4}=17000$ K, $a=0.24$,
$n=0.95$, $\alpha=0.0$, and $\bar{F}=0.7$, with typical instrumental
effects, are in Figs. \ref{F:res}, \ref{F:resf}, and \ref{F:resa},
respectively.  The filters are noted on the plot.
As shown in Fig. \ref{F:res}, box size and resolution differences are
generally within tolerance for $P$ and $B$ with the higher filter shown.
Any small box size effects
may be attributed to the difference in random
seed in the 80 $h^{-1}$Mpc boxes versus the 40 $h^{-1}$Mpc boxes.
There is a noticeable difference for $B$, the lower filter, between the
higher and lower resolution 40 $h^{-1}$Mpc boxes.  
This difference
cannot be attributed to a seed difference since the seed is same for
these boxes.  The difference is not evident between the 40 and 80
$h^{-1}$Mpc 512$^3$ boxes, because of either a box-size effect or
random fluctuations that oppose this resolution effect.  
For the most part, the $k$-dependent shapes of the derivatives agree,
with the difference being one of magnitude.
A smaller version of the Fisher analysis done in section
\ref{S:fisher} was completed for these boxes of different resolutions
to check the effects of the derivative differences.  
The errors on amplitude are about 13\% higher and on flux are about
15\% lower with the lower resolution box than with the higher
resolution box.  However, since we are only looking at error bounds,
which should hopefully be within about 25\% accuracy, this error is
tolerable, and the difference in Figure~\ref{F:res} is (by this
criterion) not significant.

\subsection{Random fluctuations due to finite box size}
\begin{figure}    
\centerline{\psfig{file=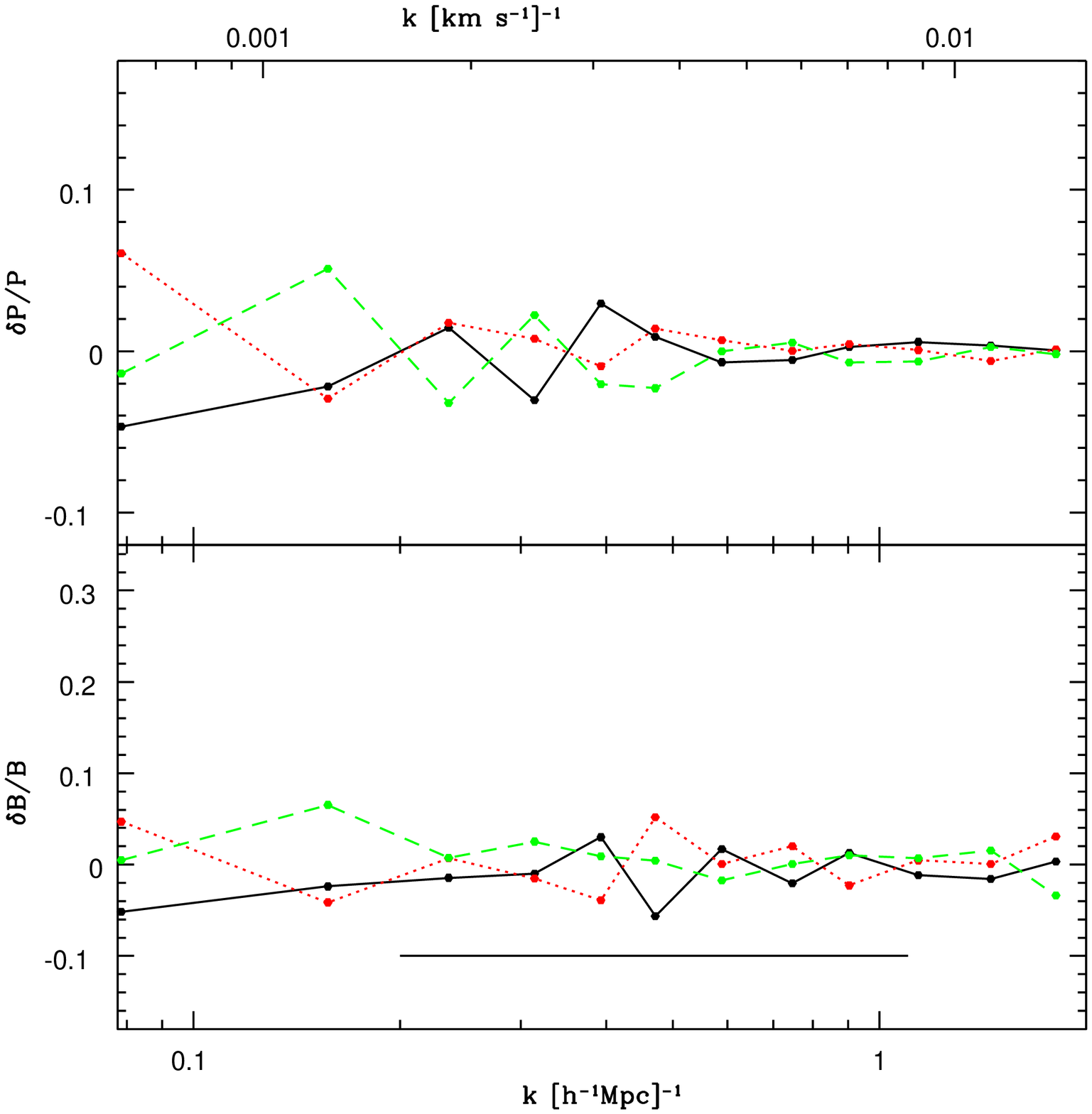,width=3.5in}}    
\caption{Relative difference between the values of the statistics
computed with three different seeds and the average value
(filter shown on plot) in an 80~$h^{-1}$Mpc box.}
\label{F:seed}    
\end{figure}

\begin{figure}     
\centerline{\psfig{file=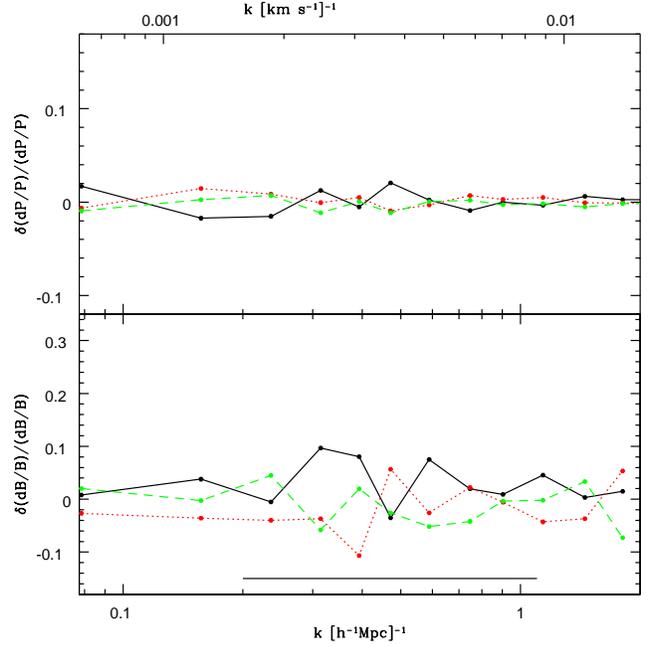,width=3.5in}}     
\caption{Relative difference between the values of $\bar{F}$-derivatives 
($\delta \bar{F}$=0.02) computed with three different seeds and the averaged
spectra (filter shown on plot) in
an 80 $h^{-1}$Mpc box.}
\label{F:seedderiv}     
\end{figure}

\begin{figure}     
\centerline{\psfig{file=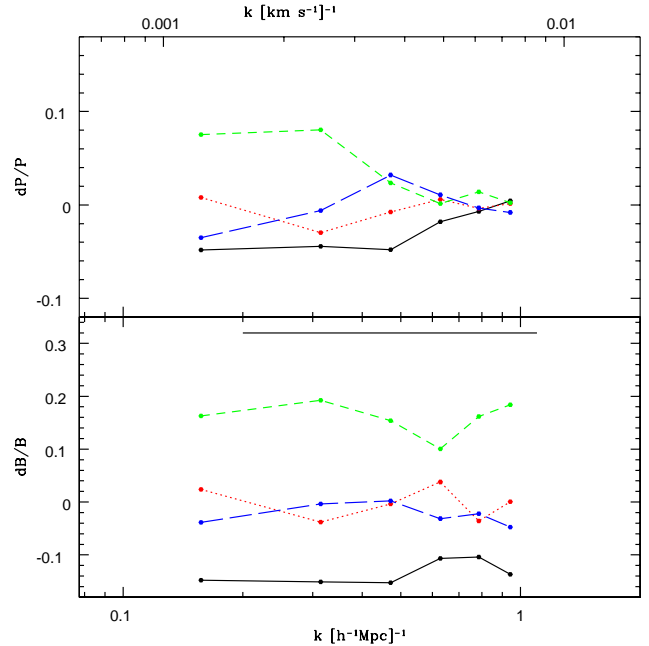,width=3.5in}}     
\caption{Relative difference between the values of the statistics
computed with four different seeds and the average value (filter shown
on plot) for a 40 $h^{-1}$Mpc box.  The reason for the systematic
difference between the $B$ values for different seeds is that these
seeds gave higher or lower than average values of $P$ for several $k$,
and since those $k$ were within the filter, that difference was 
spread systematically to all $k$.}
\label{F:seed40}     
\end{figure}

Another question is the magnitude of the random fluctuations due to
finite box size, which can be determined by using different random seeds
to generate the initial random Gaussian field.
A study was done using three seeds, with $\gamma-1=0.3$, $T_{1.4}=17000$ K, $a=0.24$,
$n=0.95$, $\alpha=0.0$, $\bar{F}=0.68$, and standard
resolution and noise effects, on the values of $P$, $B$, and
their $\bar{F}$- and $\kappa$-derivatives.
Plots of the relative difference between the statistics and their
$\bar{F}$-derivatives computed with individual seeds and averaged over
three seeds are in Figs. \ref{F:seed} and \ref{F:seedderiv}
respectively.  A plot of the relative difference between the
statistics computed with individual seeds and averaged over four seeds
for a 40~$h^{-1}$Mpc box is in Fig.~\ref{F:seed40}.

As shown in these figures, the statistics and their derivatives for an
80 $h^{-1}$Mpc box do not have very large fluctuations due to finite
box size, so only one random seed (the same each time, for
consistency) was
used for this analysis rather than averaging over the statistics
obtained from several seeds.  To check concretely that this is not a
problem, a Fisher analysis for just $\kappa$ (derivatives not shown,
but they have a similar pattern as the flux derivatives) and $\bar{F}$ at one $z$ was completed
using three different seeds, and using the spectra and covariance
matrices computed by averaging over lines of sight from all three
seeds.  The $\bar{F}$ and $\kappa$ error bounds for individual seeds were within
0.015 (relative difference) of the bounds for the averaged spectra.
Note that the 40 $h^{-1}$Mpc box has significantly
higher fluctuations, as expected, so it is possible that
apparent box size effects are actually due to these fluctuations.

\subsection{Chunking}
The simulations were also used to study the effects of chunking, a
procedure often used for data analysis.  In the data,
the spectra can be much longer than 80 $h^{-1}$Mpc, and are cut into smaller chunks for analysis.  The fields $\delta_{F}(\lambda)$
and $H(\lambda)$ are both created from the full spectrum in real space
before chunking.  

\begin{figure}
\centerline{\psfig{file=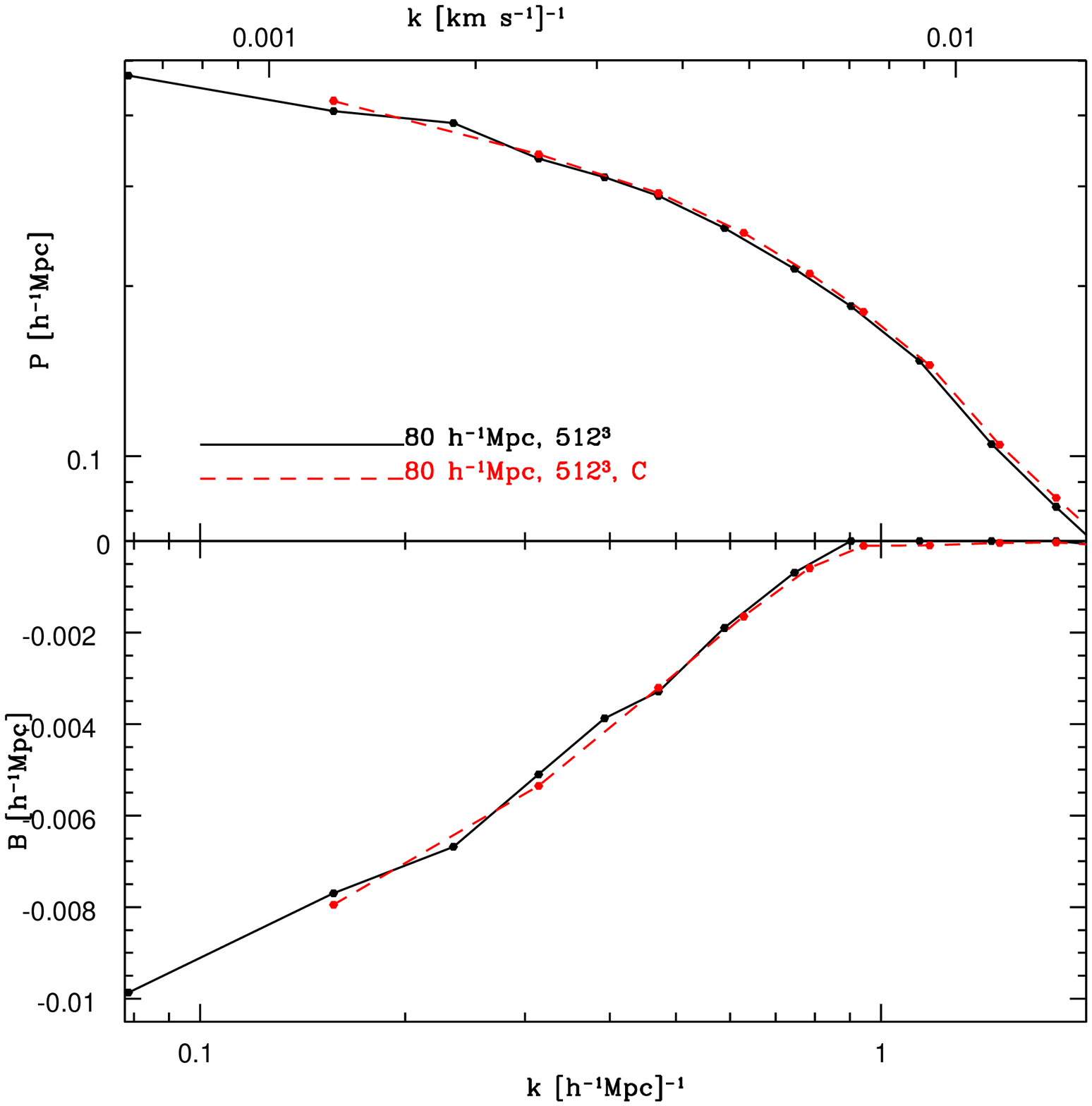,width=3.5in}}
\caption{Plot of the statistics using full lines of sight
from an 80 $h^{-1}$Mpc simulation, and a
chunked 80 $h^{-1}$Mpc simulation.  The
code ``C'' refers to chunking.}  
\label{F:periodicity}
\end{figure}

To identify the effects of 
chunking, we compared the values of the statistics for an 80
$h^{-1}$Mpc box with $512^3$ particles; and the same 80 $h^{-1}$Mpc box, computed with
half-lines of sight and averaged (see results in Fig.~\ref{F:periodicity}).
All other simulation parameters were identical: $\gamma-1 =
0.3$, $T_{1.4}=17000$ K, $a=0.24$, $n=0.95$,
$\alpha=0.0$, $\bar{F}=0.7$, and typical noise and resolution effects.

As shown in Fig.~\ref{F:periodicity}, chunking has
little effect on $P$ and $B$, with the values of the statistics
computed using chunking almost perfectly matching those for the 80 $h^{-1}$Mpc box
without chunking.  Thus, the chunks of spectrum as small as 40 $h^{-1}$Mpc used in
a data analysis can accurately represent the results in an
infinite-spectrum limit.

\subsection{Filters}\label{S:filters}

As stated in section \ref{S:statistics}, we tried two filter shapes and several filter sizes and numbers of filters.
The square window, in Eq.~\ref{E:SquareW}, was used in \citet{2001ApJ...551...48Z}.  
The Gaussian window was also tried since the Fourier transform of
the square filter is not well-localized.  
The sizes of the filters were chosen based on the following
consideration: the $k$ values in these 80 $h^{-1}$Mpc boxes range from 0.078-5.0~$h$Mpc$^{-1}$.  
Assuming $\Omega_m=0.3$, at $z=3$, this range corresponds to 0.0007-0.05 s km$^{-1}$.  
Below about 0.001 s km$^{-1}$, continuum fluctuations become
important, and above about 0.02 s km$^{-1}$, resolution and noise
effects become very important. 
Consequently, for square filters, the largest range of filter used is
0.002-0.02 s km$^{-1}$ (0.2-2.0 $h$Mpc$^{-1}$), and several smaller filters were chosen.  
For Gaussian filters, the question of filter width is a bit more
difficult because there is some small contribution from beyond even
$2\sigma$.  Thus, we hoped to find the balance between the competing
considerations of filter smoothness (compactness of the Fourier
transform) versus its ability to eliminate
$k$-values outside the range we want to consider.  For the data, other possibly
useful filters that can balance the smoothness and noise
considerations are a triangular filter, or a Gaussian filter with a
sharper low-$k$ cutoff.
We show results for several sizes of square and Gaussian filters.

First we investigated how does an increase in number of filters with 
different sizes improve the information from $B$.
A Fisher analysis using varying numbers of square filters was done to determine the number of filters to use in the range 0.2-2.0 $h$Mpc$^{-1}$.
We tried using one or two filters by subdividing that interval accordingly.  
No filters overlapped, as analysis showed that overlap did not increase the amount of information at all.  
The error bars computed from one filter versus from two filters in the
same range were within 5\% of each other.  Thus one filter is
sufficient for analysis of a given range of $k$.  However, for this
analysis we used two non-overlapping filters for the total range, allowing us to
analyze separately the information contribution of lower and higher
filter ranges.

We also investigated filter sizes and shapes as shown
in Table \ref{T:filters}.
Filter B1 is the lower filter of the pair referred to as filter B, and B2 is the higher filter.  
The numbers shown are actually for $z=3$; the filter parameters are
constant in s km$^{-1}$, and so their value in $h$Mpc$^{-1}$ is
$z$-dependent.  While the magnitude of the statistics are
rather filter-dependent, the shape of the statistics is
essentially filter-independent and so the results were 
almost independent of the filter size and shape used, at least 
within the range of filters explored here.  

\begin{table}
\begin{center}
\begin{tabular}{|| c | c c | c c||}
\hline
 & \multicolumn{2}{|c}{Filter 1} & \multicolumn{2}{c}{Filter 2} \\ \hline
 & \multicolumn{4}{|c}{Square Filters} \\ 
 & Start & End & Start & End \\ \hline
B & 0.2 & 1.1 & 1.1 & 2.0 \\
C & 0.4 & 1.2 & 1.2 & 2.0 \\
D & 0.2 & 1.0 & 1.0 & 1.8 \\ 
E & 0.4 & 1.1 & 1.1 & 1.8 \\ \hline
 & \multicolumn{4}{|c}{Gaussian Filters} \\ 
 & $\bar{k}$ & $\sigma_k$ & $\bar{k}$ & $\sigma_k$ \\ \hline
F & 0.65 & 0.2 & 1.5 & 0.3 \\ 
G & 0.65 & 0.13 & 1.5 & 0.2 \\ 
H & 0.65 & 0.27 & 1.5 & 0.4 \\ 
\end{tabular}
\caption{Table of codes for the seven filters analyzed.  All figures are in $h$Mpc$^{-1}$ at $z=3$.}
\label{T:filters}
\end{center}
\end{table}

\subsection{Instrumental dependencies}\label{S:noise}
Noise and resolution play an important role in determining the sensitivity
of a given data set.
We parameterize the noise with two numbers, $f_{c}$
(the fraction of noise whose amplitude is correlated with the signal) and $S/N$ (the overall
signal to noise ratio relative to the continuum, including all noise).  Noise whose amplitude is 
correlated with the signal is characterized by its variance
$\sigma_c^2 = c\bar{F}$ (where $\bar{F}$ is the mean flux from the
quasar).
A typical example is photon shot noise, where the variance 
is proportional to the flux. Noise whose amplitude is uncorrelated with the 
signal is characterized by
its variance $\sigma_u^2$ (for example, Gaussian readout noise and 
photon shot noise from the sky)
These variables are related as follows, defining a unique $c$
and $\sigma_u^2$ for each $f_{c}$ and $S/N$:
\begin{align}\label{E:noisedef}
f_{c} &= \frac{c\bar{F}}{c\bar{F}+\sigma_u^2}\\ 
S/N &= \frac{1}{\sigma_f} = \frac{1}{\sqrt{\sigma_u^2+c\bar{F}}}
\end{align}

The distinction between correlated and uncorrelated noise amplitude is made 
because they have different effects on $B$ and $T$.  
Uncorrelated noise contributes to $P$ and $T$, but not to $B$,
whereas correlated noise contributes to all three.  The noise's contribution to
$P$ can be calculated analytically whether its amplitude is 
correlated or uncorrelated with the signal. The correlated noise 
contribution to $B$ depends not only on the noise amplitude, but also on the 
signal power spectrum $P$.  

In this study, we tried to match the pixelization, resolution, and noise to that of the SDSS data.  
The pixels in SDSS data are roughly 1 \AA{} wide
(so the smoothing $\sigma$ is approximately the size of one
pixel).  
Assuming $z=3$ and $\Omega_m=0.3$ this corresponds to 0.6 $h^{-1}$Mpc.  
To match the 0.6 $h^{-1}$Mpc pixels, we used pixels 
of size 0.625 $h^{-1}$Mpc in the simulations.  This small difference from
the real value makes
little effect in the value of the statistics, since even reducing the
pixel size by a factor of four changes  $B$ and $P$ by at most 5\%.
We used Gaussian smoothing $\sigma$ the same size as our pixels.  
SDSS data have
$S/N$ (averaged over redshift bins in our range) ranging from 1 to
30, with median values in each redshift bin ranging from 4 to 5.
We used the value $S/N =5.0$ for all redshift bins.  The noise amplitude was entirely uncorrelated with the signal ($f_c=0$), since we could not evaluate the amount of correlated noise in the data. 

\section{Fisher Matrix Calculation and Results}\label{S:fisher}

%%%Moving this paragraph to the beginning of simulations section as referee suggested.
%To assess the sensitivity of the Ly-$\alpha$ forest to cosmological parameters
%we performed a Fisher matrix analysis (e.g. \citet{1997ApJ...480...22T}).
%We conducted a  redshift-dependent analysis 
%using 9 redshift bins, centered at $z=2.2, 2.4, \ldots, 3.8$ and using 
%a realistic redshift distribution of quasars as found in SDSS data.  
%Since we only had simulations with values of $a$ spaced by 0.04, we chose the closest $a$ available, and used a value of $\bar{F}$ from the fit in Eq. \ref{E:Fz}.
%
The construction of the Fisher matrices for each individual redshift bin is straightforward.  
A vector containing all values $x_i$ that were to be used (for example, $P$ and $B$ for all $k$) was created, as was its full covariance matrix (computed using all 12288 lines of sight) and its derivatives with respect to the 6 parameters 
$\gamma$, $T_{1.4}$, $\kappa$, $n$, $\alpha$, and $\bar{F}$.  
  The derivatives with respect to $P$ were created from the values with the 
power due to noise subtracted off.
These covariance matrices and derivatives were then used to create the $6\times6$ Fisher matrices for each redshift bin by computing
\begin{equation}\label{E:fishdef}
F_{kl} = \left(\frac{\mathrm{d}\mathbf{x}}{\mathrm{d}y_k}\right)^T \cdot \mathsf{C}^{-1} \cdot \left(\frac{\mathrm{d}\mathbf{x}}{\mathrm{d}y_l}\right)
\end{equation}
so that $\mathsf{F}$ is the inverse covariance matrix for $\mathbf{y}$, the vector of all the parameters.
In general, there is another term related to
$\mathrm{d}\mathsf{C}/\mathrm{d}y_i$ that we neglect in the expectation that it is small.  

%%%Also moved this to part of simulations section since it is an
%%%explanation of the derivatives.
%Instead of varying the actual power spectrum amplitude, which requires a large number of different simulations, we varied $a$.  
%In relating derivatives such as $\mathrm{d}P/\mathrm{d}a$ to the
%desired derivatives $\mathrm{d}P/\mathrm{d}\kappa$, we used the
%approximation (shown in \citet{2001astro.ph..8064M} to be quite
%accurate) that $z$-evolution of the power spectrum can be treated as a
%rescaling of $\kappa$ and $T_{1.4}$.  The change in growth factor can
%be treated primarily as a change in $\kappa$, along with a small change in $T_{1.4}$ (fixed in velocity
%coordinates, it consequently differs in
%simulation coordinates).
%Thus, by subtracting off this temperature term from the
%$a$-derivatives, we obtain the portion of the $a$-derivative that
%mimics a change in power spectrum amplitude evolution.  Then we use
%$\mathrm{d}\kappa = \mathrm{d}a/a$ to form the amplitude derivative
%(this form assumes an Einstein-de Sitter universe with $D(a)=a$).

Next, a block-diagonal $54 \times 54$ Fisher matrix was created, using
the $6\times6$ Fisher matrices for the nine redshift bins.  
Each block was multiplied by the length of spectrum in that redshift
bin in the SDSS data; 
%%%Added the rest of this sentence in response to referee question.
the total spectrum length is $7.7\times 10^7$ km/s.
Fisher matrices in this form were created for all seven sets of
filters listed in Table \ref{T:filters}, for various combinations of
statistics.  Further information about error-bounds is extracted from these large Fisher matrices by projecting from them down to a smaller number of parameters.

\subsection{Projection to fewer parameters}\label{SS:projection}

To reduce to the smallest number of parameters, we used the following procedure.  The $54\times54$ Fisher matrices are the inverse covariance 
matrices $\mathsf{C}^{-1}(\mathbf{y})$ for the values of the parameters in each redshift bin.  
We projected down to a smaller set of parameters that includes:
\begin{itemize}
\item $\gamma_0$ and $\gamma_1$, the constant and linear terms of a linear expansion of $\gamma(z)$ about $z=3.0$, 
\item $T_{1.4}$, an overall redshift-independent value, 
\item $\kappaeff$, an overall effective amplitude at some $z$ and $k$,
\item $\neff$, an overall effective slope, where the evolution of the pivot point (fixed in velocity
coordinates) is expressed via $n(z)=n(k_{p})+\alpha\ln{(k_{p}(z)/k_{p})}$,
\item $\alphaeff$, an overall effective curvature, and
\item The nine values of $\bar{F}$, one per redshift bin.
\end{itemize}
This projection includes 15 parameters, and can be computed using Eq.~\ref{E:fishdef}. 

Here we present results for the analysis done with just filter B, using $P$ and $B$.  
All references to ``error bounds'' implies
that we have marginalized over other parameters (by using
$\sqrt{(F^{-1})_{ii}}$ rather than $1/\sqrt{F_{ii}}$).

First, the error bounds for $\gamma_0$ and $\gamma_1$, defined by
[$\gamma-1](z) = \gamma_0 + \gamma_1 (z-3)$, are 0.054 and 0.10,
respectively.  
We are not very sensitive to $T_{1.4}$, with a relative error of 0.08.

Table \ref{T:fluxerr} shows the relative errors for the nine values of
$\bar{F}$. The errors increase with redshift mainly because the amount
of data decreases, and at low redshifts they are extremely small.
\begin{table}
\begin{center}
\begin{tabular}{| c | c | c |}
$z$ & $\bar{F}$ & $\Delta \bar{F}/\bar{F}$ \\ \hline
2.2 & 0.85 & 0.0030 \\
2.4 & 0.81 & 0.0035 \\
2.6 & 0.77 & 0.0033 \\
2.8 & 0.73 & 0.0036 \\
3.0 & 0.68 & 0.0046 \\
3.2 & 0.63 & 0.0057 \\
3.4 & 0.57 & 0.0075 \\
3.6 & 0.52 & 0.011 \\
3.8 & 0.46 & 0.015 \\ 
\end{tabular}
\caption{Error bounds on $\bar{F}(z)$ using $P$ and $B$.}
\label{T:fluxerr}
\end{center}
\end{table}

\begin{figure}
\centerline{\psfig{file=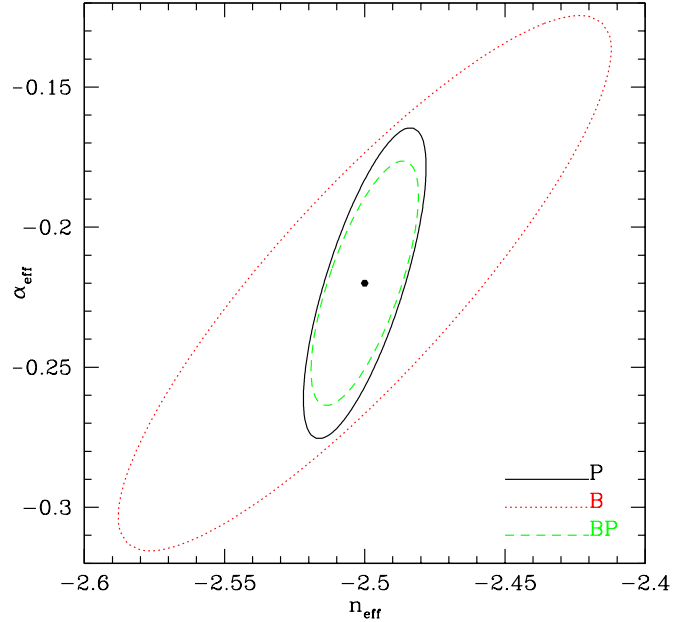,width=3.5in}}
\caption{68\% CL error ellipses for $\neff$ and $\alphaeff$ computed with various combinations of statistics.}
\label{F:nalpha_b_stat}
\end{figure}

The error on $\kappa=(\Delta \sqrt{A})/\sqrt{A} $ is .017.  
The errors on $\neff$ and $\alphaeff$ (defined at pivot point $k=0.009$ s km$^{-1}$ at $z=2.6$) are $\Delta \neff = 0.013$, and $\Delta \alphaeff = 0.029$.  
The error ellipse in terms of $\neff$ and $\alphaeff$ computed from their joint covariance matrix is in Fig. \ref{F:nalpha_b_stat}.
To answer the question of whether $B$ alone provides independent 
information competitive wih $P$, 
or whether $B$ and $P$ together is more sensitive 
than either of the two alone, we
carried out the analysis using $P$
alone, $B$ alone, and $P$ and $B$ together.
The error bounds for all parameters for various combinations
of statistics can be found in Table \ref{T:ampstat}.  
\begin{table}
\begin{center}
\begin{tabular}{| c | c | c | c |}
 & $P$ & $B$ & $PB$ \\ \hline
$\Delta \gamma_0$ & 0.097 & 0.086 & 0.054 \\
$\Delta \gamma_1$ & 0.18 & 0.17 & 0.10\\
$\Delta T_{1.4}/T_{1.4}$ & 0.096 & 0.093 & 0.081 \\
$\Delta \kappaeff$ & 0.045 & 0.035 & 0.017 \\
$\Delta \neff$ & 0.014 & 0.058 & 0.013 \\
$\Delta \alphaeff$ & 0.037 & 0.063 & 0.029 \\
$\Delta \bar{F}_1/\bar{F}_1$ & 0.0082 & 0.0057 & 0.0030\\ 
\end{tabular}
\caption{Errors using various combinations of statistics
(note that $\bar{F}_1$ is the mean flux for the first redshift bin,
and illustrates the trend in errorbars for all mean fluxes).}
\label{T:ampstat}
\end{center}
\end{table}

Fig. \ref{F:nalpha_b_stat} shows the 68\% confidence level error ellipses
for $\neff$ and $\alphaeff$ for $P$, $B$, and both together.
It is clear that for this combination of parameters $P$ provides most of 
the information. While $B$ can act as a useful independent confirmation,
adding its information by itself does not decrease the statistical 
errors significantly. 

\begin{figure}
\centerline{\psfig{file=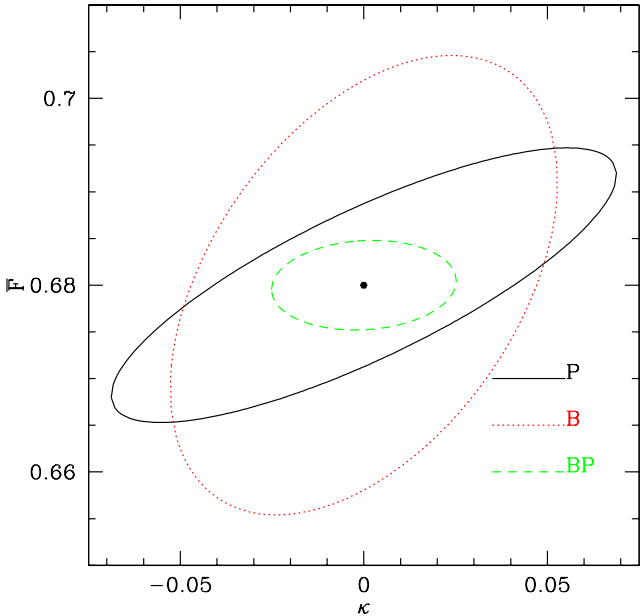,width=3.5in}}
\caption{68\% CL error ellipses for $\kappaeff$ and $\bar{F}(z=3)$ computed with various combinations of statistics.  
Note that the correlation coefficient between these parameters 
is 0.80 for $P$, 0.45 for $B$, and 0.10 for $B$ and $P$ together.  
The reason that the correlation is so reduced when $B$ and $P$ 
are used together is that there are nonzero covariance matrix 
elements between $B$ and $P$ at the same $k$, so the Fisher matrix 
is not that produced by simply adding those for $B$ and $P$ together.}
\label{F:amp_f_stat}
\end{figure}

Fig.~\ref{F:amp_f_stat} shows the 68\% CL error ellipses for $\kappaeff$ and $\bar{F}(z=3.0)$ for $P$, $B$, and both together. 
In contrast with Fig.~\ref{F:nalpha_b_stat}, combining the two  significantly
improves upon $P$ or $B$ alone, reducing the errors on $\kappa$ and $\bar{F}$
by almost a factor of 3. The primary reason for this is the partial degeneracy 
between the two parameters in $P$ alone, as shown in Fig.~\ref{F:amp_f_stat}.
This degeneracy is easy to understand from the derivatives in Figs.~\ref{F:powparams} and~\ref{F:otherparams}.
Changing $\kappa$ or $\bar{F}$ has a relatively 
flat effect on the power spectrum $P$, so the two parameters are 
significantly degenerate from $P$ alone, especially when all the other parameters
are included. 
In contrast, the effect on $B$ is much larger for $\kappa$ than for $\bar{F}$.
Combining $P$ and $B$ thus allows one to break the degeneracy between the 
two parameters and determine both with a much higher precision. 


We also investigated how the errors vary with filter, finding that 
they did not vary significantly for filters given in Table \ref{T:filters}. 
In particular, the error bounds on $\neff$ and $\alphaeff$ were virtually
independent of filter, and those on $\kappaeff$ were nearly so.  Among
the Gaussian filters, there was a slight trend towards lower error
bounds on $\kappaeff$ for the larger filters (0.015 for the largest
versus 0.018 for the smallest) but even here the difference is not
very large.  
%The error bounds on $\gamma-1$ and $T_{1.4}$ did not vary greatly with filter,
%ranging from 0.051-0.064 for $\gamma_0$, 0.10-0.12 for
%$\gamma_1$, and 0.080-0.085 relative error on  $T_{1.4}$.  
%The flux error bounds likewise did not vary significantly with
%filter.  
%The bounds on amplitude were quite similar for all filters, ranging
%between 0.015 and 0.018 relative error, as were the error ellipses for $n$ and $\alpha$.  
%A comparison of the error
%ellipses for $n$ and $\alpha$ is in Fig. \ref{F:nalphafilt}.
%\begin{figure}
%\centerline{\psfig{file=fig16.ps,width=3.5in}}
%\caption{$2\sigma$ error ellipses for $n$ and $\alpha$ computed using
%$P$ and $B$, for all filters.}
%\label{F:nalphafilt}
%\end{figure}

\subsection{Expansions of $\kappaeff$, $\neff$, and $\alphaeff$}\label{SS:expansions}

In order to determine how much variation in the values of $\neff$,
$\alphaeff$, and $\kappaeff$ (at a particular $z$ and $k$) is induced
by deviations in $n$, $\alpha$, $\Omega_m$, $\Omega_b$, and $h$, we
used CMBFAST \citep{1996ApJ...469..437S} 
outputs to study the derivatives with respect to these
parameters, using spectra that were normalized to the same amplitude
at $k=0.05$ $h$Mpc$^{-1}$ at $z=20$.  
The derivatives were computed at
$z=2.6$, $k=0.009$ s km$^{-1}$, the pivot point for SDSS data sample. 
We assume the fiducial model 
$\Omega_m=0.3$, $h=0.7$, $\Omega_b=0.04$ (or $x\equiv \Omega_b/\Omega_m=
0.13$).  The pivot point for $\alpha$ is $k=0.05$ $h$Mpc$^{-1}$.  The
expansion for $\kappaeff$ is denoted $\Delta \kappaeff$ to emphasize
that this expansion is to be used to calculate what relative change in
power spectrum amplitude would result from changing the cosmological
parameters from our chosen values (hence it gives $\Delta \kappaeff=0$
for our cosmological model).
The resulting expansions are
\begin{align}
\Delta \kappaeff &= \Delta \kappa+1.5(n-0.75) - 0.032\alpha \\ \notag
 & \quad +0.28\ln{\left(\frac{\Omega_m}{0.3}\right)}
 +0.65\ln{\left(\frac{h}{0.7}\right)} + 0.14\ln{\left(\frac{x}{0.13}\right)} \\
\neff &= -3.25 + n + 3.0\alpha + \\ \notag 
 & \quad 0.16\ln{\left(\frac{\Omega_m}{0.3}\right)}
 +0.23\ln{\left(\frac{h}{0.7}\right)} +0.034\ln{\left(\frac{x}{0.13}\right)} \\
\alphaeff &= -0.22 + \alpha
-0.11\ln{\left(\frac{\Omega_m}{0.3}\right)} \\ \notag
 & \quad -0.17\ln{\left(\frac{h}{0.7}\right)}
 + 0.034\ln{\left(\frac{x}{0.13}\right)} 
\end{align}

The effect on the amplitude is primarily determined by the 
primordial amplitude and slope, but not the derivative of the slope (given that the pivot point $k=0.05$ $h$Mpc$^{-1}$ is at much lower $k$ than the wavevectors the forest is sensitive to). 
Today's matter density
$\Omega_m$ and Hubble constant $h=H_0/100$ ${\rm km/s/Mpc}$ change the transfer function and so affect 
the amplitude, slope and derivative of the slope. 
However, $\Omega_m$ also 
changes the relation between velocities and comoving coordinates as given 
by the Hubble parameter $H(z)$. This partially cancels the transfer 
function effect of $\Omega_m$, so that
the coefficients in front of $\Omega_m$ are reduced relative to $h$. 
Baryon density has  a smaller effect than other parameters, at least
around the small value of $\Omega_b/\Omega_m$ we expanded here. 

It is clear that since we can only determine precisely two numbers, 
the amplitude and the slope, and somewhat less precisely also the 
derivative of the slope, we cannot determine all of the parameters 
with high precision. Hence it is best if the Ly-$\alpha$ forest is
combined with other cosmological probes, such as cosmic 
microwave background (CMB) anisotropies.
Since one expects 
CMB to provide very strong constraints on $h$ and $\Omega_m$ (at 
least in the context of spatially flat models), this then allows 
the Ly-$\alpha$ forest to determine the primordial slope and its
derivative with high accuracy.

\section{Power Spectrum Evolution and Quintessence}\label{S:quintessence}

In this section we try to determine the sensitivity of the Lyman-$\alpha$
forest to deviations from the Einstein-de Sitter (EdS) growth factor
$D(a)=a$. We analyzed both specific quintessence models and also more 
general deviations of the growth factor from EdS.  
Sensitivity to quintessence model parameters can 
arise from the change in the growth factor $D(z)$ and from the 
Hubble parameter $H(z)$.
In our tests the latter was subdominant and most of the 
sensitivity was from the growth factor. 

First, we studied the sensitivity of the Lyman-$\alpha$
forest to quintessence parameters, $\Omega_q$ and the equation of
state $w_q$, for nine different quintessence models.  Several
models considered had static equation of state $w_q = w_0$, while
 the other models were dynamic,
$w_q = w_0 + w_1 (a-1)$.  The models studied are shown in Table
\ref{T:qmodels}.  
\begin{table}
\begin{center}
\begin{tabular}{| c | c | c | c | c |}
Model & $\Omega_{q,0}$ & $\Omega_q(z=2.6)$ & $w_0$ & $w_1$ \\ \hline
1 & 0.67 & 0.12 & -0.7 & 0.0 \\ 
2 & 0.85 & 0.28 & -0.7 & 0.0 \\ 
3 & 0.49 & 0.06 & -0.7 & 0.0 \\ 
4 & 0.67 & 0.04 & -1.0 & 0.0 \\ 
5 & 0.67 & 0.30 & -0.4 & 0.0 \\ 
6 & 0.67 & 0.12 & -0.8 & -0.2 \\ 
7 & 0.67 & 0.19 & -0.8 & -0.5 \\ 
8 & 0.67 & 0.27 & -0.8 & -0.8 \\ 
9 & 0.67 & 0.09 & -1.0 & -0.5 \\ 
\end{tabular}
\caption{Codes for quintessence models, where $w_q=w_0+w_1 (a-1)$.}
\label{T:qmodels}
\end{center}
\end{table}
All models had $\Omega_b = 0.04$ and $h = 0.7$, with $\Omega_{\mathrm{CDM}}$
determined by requiring $\Omega_b + \Omega_{\mathrm{CDM}} + \Omega_q = 1$.

\subsection{Projection with quintessence parameters}\label{SS:results}

To find constraints on all parameters once quintessence parameters are
included, we started from the $54\times 54$ Fisher matrices created in
section \ref{S:fisher}, and projected down to the parameters in
section \ref{SS:projection} plus $\Omega_q$ and $w_q$ (or $\Omega_q$, $w_0$, and
$w_1$ for dynamical models).  All error bounds are computed from
the Fisher matrix for filter B using $P$ and $B$.
To do so, we computed derivatives of $\kappaeff$, $\neff$, and $\alphaeff$ with respect to quintessence
parameters at some pivot point fixed in s km$^{-1}$.
These derivatives were computed from CMBFAST outputs
for the 9 $z$-bins for all 9 quintessence
models studied, normalized to have the same amplitude for $z=20$ at
the same pivot point at which derivatives were taken.

Results for all models indicated that $w_q$ and $\Omega_q$ are highly
degenerate, with a correlation coefficient of -0.9.  
The well-determined direction for
model 1, which is fairly representative, in $(w_q,\Omega_q)$ space is $(0.60,0.80)$.
The strong correlation is not surprising,
since the redshift range probed is too small to determine two parameters
from the growth factor or $H(z)$ evolution. Some information on $\Omega_q$
also comes from the power spectrum shape, which is why the correlation is 
not perfect.
In the following we thus fix one parameter, for example
$\Omega_q$, and present errors on equation of state only.
The motivation for this is that
other tests, most notably CMB combined with large scale structure tests
(e.g. cluster counts, galaxy clustering, weak lensing),
will be able to determine $\Omega_q$ at $z=0$ very accurately.
If these tests also provide independent constraints on $w_0$
then for the time dependent $w$ one can use our results to constrain $w_1$.

\begin{figure}
\centerline{\psfig{file=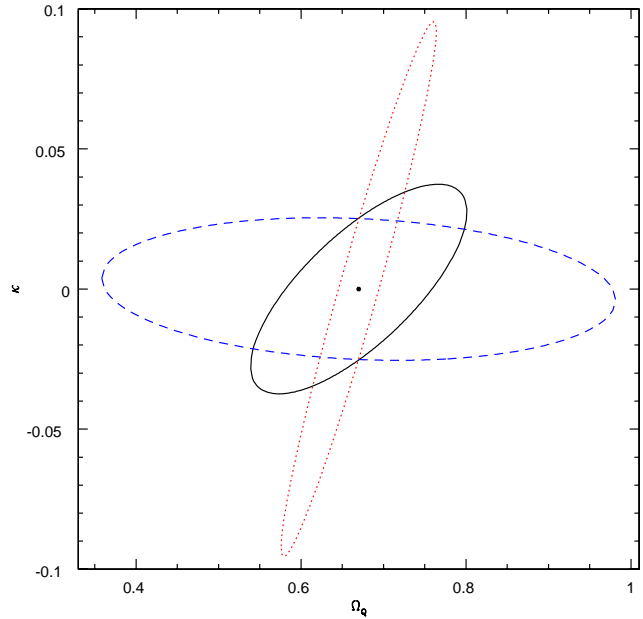,width=3.5in}}
\caption{68\% CL error ellipse for $\Omega_q$ and amplitude, with
$w_q$ (but not $\alphaeff$) fixed, for cosmological constant (model 4)
as dashed lines, quintessence with $w=-0.4$ (model 5) as solid lines,
and a dynamic quintessence model (model 8) as dotted lines.}
\label{F:Omegaamp}
\end{figure}

The errors for fixed $w_q$ (marginalized over all other variables) 
can be found in Table \ref{T:qfixedw}.
The 68\% CL error ellipses show how $\Omega_q$ and amplitude are
degenerate (Fig. \ref{F:Omegaamp}).  Analysis of the covariance matrices
for these model parameters (or the fact that the error bounds for
$\kappaeff$, $\neff$, and $\alphaeff$ have risen from their value
computed without including quintessence) indicated a very high degree of degeneracy
between $\Omega_q$ and $\kappaeff$, $\neff$, and $\alphaeff$.  If
$\alpha$ is fixed then the errors on $\Omega_q$
consequently decrease (though this effect is much greater for the
models to which we are least sensitive).  This indicates that at least some of our
information about quintessence parameters comes from shape
information.  The degeneracy with $\kappaeff$ indicates that some
information is from the growth factor information as well, since the
magnitude of the growth factor (as opposed to its shape) is degenerate
with $\kappaeff$.  The errors with $w$ and $\alpha$ fixed are in Table~\ref{T:qfixedwalpha}.
\begin{table}
\begin{center}
\begin{tabular}{| c | c | c | c | c |}
Model & $\Delta \Omega_q/\Omega_q$ & $\Delta \kappaeff$ &
$\Delta \neff$ & $\Delta \alphaeff$ \\ \hline
1 & 0.26 & 0.041 & 0.057 & 0.052 \\
2 & 0.05 & 0.073 & 0.047 & 0.034 \\
3 & 0.48 & 0.028 & 0.075 & 0.053 \\
4 & 0.31 & 0.017 & 0.097 & 0.078 \\
5 & 0.13 & 0.025 & 0.049 & 0.038 \\
6 & 0.16 & 0.037 & 0.053 & 0.049 \\
7 & 0.11 & 0.048 & 0.040 & 0.039 \\
8 & 0.09 & 0.063 & 0.035 & 0.033 \\
9 & 0.20 & 0.030 & 0.066 & 0.052 \\ 
\end{tabular}
\caption{Error bounds for all quintessence models with fixed equation
of state, marginalized over
other parameters.}
\label{T:qfixedw}
\end{center}
\end{table}
\begin{table}
\begin{center}
\begin{tabular}{| c | c | c | c |}
Model & $\Delta \Omega_q/\Omega_q$ & $\Delta \kappaeff$ &
$\Delta \neff$ \\ \hline
1 & 0.14 & 0.024 & 0.039 \\
2 & 0.04 & 0.059 & 0.043 \\
3 & 0.26 & 0.018 & 0.049 \\
4 & 0.11 & 0.017 & 0.043 \\
5 & 0.10 & 0.019 & 0.041 \\
6 & 0.09 & 0.022 & 0.037 \\
7 & 0.08 & 0.034 & 0.034 \\
8 & 0.08 & 0.051 & 0.033 \\
9 & 0.11 & 0.019 & 0.044 \\ 
\end{tabular}
\caption{Error bounds for all quintessence models with fixed equation
of state and $\alpha$, marginalized over
other parameters.}
\label{T:qfixedwalpha}
\end{center}
\end{table}

Error bounds on $w_q$ for static models, with $\Omega_q$ fixed (as may be done
by measurements other than the Lyman-$\alpha$ forest), are shown in
Table~\ref{T:qfixedomega}.  As shown there, now that $\Omega_q$ is fixed, $\neff$
and $\alphaeff$ have error bounds roughly equal to those computed in
the previous section with no quintessence.  This fact shows that these
parameters contain information about $\Omega_q$ but not $w_q$.
$\kappaeff$, on the other hand, is degenerate with $w_q$, indicating
that $w_q$ is detected via its influence on the growth factor.
Consequently, the model with least negative $w_q$ has the best error
bounds.  
Table~\ref{T:qfixedomegaw0} shows error bounds
for $w_1$ with both $\Omega_q$ and $w_0$ fixed.  Errors for $\neff$
and $\alphaeff$ are not shown since they are the same as in the
previous section, without quintessence.  Not surprisingly, the error bounds are best for the
model with the most rapidly-varying equation of state.
\begin{table}
\begin{center}
\begin{tabular}{| c | c | c | c | c |}
Model & $\Delta w_q$ & $\Delta \kappaeff$ &
$\Delta \neff$ & $\Delta \alphaeff$ \\ \hline
1 & 0.22 & 0.10 & 0.016 & 0.029 \\
2 & 0.09 & 0.10 & 0.014 & 0.028 \\
3 & 0.36 & 0.08 & 0.016 & 0.029 \\
4 & 0.36 & 0.06 & 0.013 & 0.029 \\
5 & 0.17 & 0.25 & 0.025 & 0.028 \\
\end{tabular}
\caption{Error bounds for static quintessence models with fixed $\Omega_q$, marginalized over
other parameters.}
\label{T:qfixedomega}
\end{center}
\end{table}
\begin{table}
\begin{center}
\begin{tabular}{| c | c | c |}
Model & $\Delta w_1$ & $\Delta \kappaeff$ \\ \hline
6 & 0.31 & 0.06  \\
7 & 0.22 & 0.07 \\
8 & 0.18 & 0.08  \\
9 & 0.41 & 0.06  \\
\end{tabular}
\caption{Error bounds for dynamic quintessence models with fixed
$\Omega_q$ and $w_0$, marginalized over
other parameters.}
\label{T:qfixedomegaw0}
\end{center}
\end{table}

\subsection{More general growth}\label{SS:gf}

Besides considering specific quintessence models, we also considered
more general power spectrum evolution, modeled as 
\begin{equation}
D(a)=D_0(a/a_0)^s,
\end{equation}
where $s=1$ corresponds to an Einstein-de Sitter universe.  Thus, we
can determine if our sensitivity to quintessence is due
to growth factor deviations.  
This analysis was completed by starting with the $54\times54$ Fisher
matrix for $P$ and $B$ (filter B), and projecting down to the same set of parameters as in
Section~\ref{SS:projection} along with $s$.  The
projection for $\kappa$ and $s$ used $P \propto e^{2\kappa} D^2(a)$ to relate the necessary derivatives
$\mathrm{d}P/\mathrm{d}\kappa$ and $\mathrm{d}P/\mathrm{d}s$ to the
computed values $\mathrm{d}P/\mathrm{d}a$.  Using the value of $a_0$ 
determined by requiring that amplitude and
$s$ have the minimum degeneracy, and inputting a value of $s=1$, 
yields an error $\Delta s = 0.11$.  Note
that if only information from the power spectrum $P$ is used (rather than
including $B$ as done here), the error rises to $\Delta s = 0.47$.

\begin{figure}
\centerline{\psfig{file=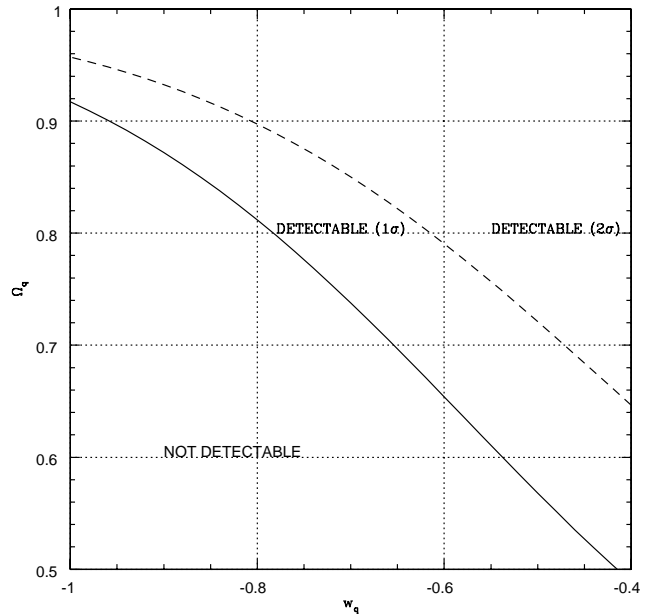,width=3.5in}}
\caption{Curve in the $\Omega_q$, $w_q$ plane indicating regions for
which the deviation from $D(a)=a$ would be detectable using growth factor information (1 and
$2\sigma$ as solid and dashed lines, respectively).  Note that the Gaussian approximation is worse for the
$2\sigma$ errors.}
\label{F:qcurve}
\end{figure}

Values of $\Delta s$ for arbitrary static quintessence models were
computed using an integral expression for $D(a)$ obtained from
\cite{2001MNRAS.322..419H}, Taylor expanding to several orders, which gave a form
$D(a)=a\left(1+\sum_{n=1} c_n(\Omega_q,w_q) a^{-3nw_q}\right)$.  Expanding this expression about $a_0$
gave an expression for $s$, and hence for $\Delta s = s-1$, in terms of the
model parameters.  Requiring $|\Delta s| < 0.11$ defines the region in
$w_q$-$\Omega_q$ parameter space that is detectable using the Lyman-$\alpha$
forest (growth factor information only, not curvature).  
This curve is shown in Fig.~\ref{F:qcurve}.

To check the validity of this expansion, the growth factor was
computed directly from CMBFAST outputs for the 9 quintessence models
studied above.  This also allowed the computation of $\Delta s$ for
dynamic models, for which an analytic expression was not obtained.
Values of $\Delta s$ for the 9 models studied earlier, and values of
$\Omega_q(z=2.6)$, are shown in Table~\ref{T:deltas}.  As shown,
$\Delta s$ and $\Omega_q(z=2.6)$ are highly correlated.
\begin{table}
\begin{center}
\begin{tabular}{| c | c | c | c |}
Model & $\Omega_q(z=2.6)$ & $\Delta s$ & Detect? \\ \hline
1 & 0.12 & -0.085 & No \\
2 & 0.28 & -0.22 & Yes \\
3 & 0.06 & -0.042 & No \\
4 & 0.04 & -0.024 & No \\
5 & 0.30 & -0.23 & Yes \\
6 & 0.12 & -0.082 & No \\
7 & 0.19 & -0.14 & Yes \\
8 & 0.27 & -0.22 & Yes \\
9 & 0.09 & -0.060 & No \\
\end{tabular}
\caption{Values of $\Delta s$ for the nine quintessence models, where
detectability of growth factor deviation is defined by $|\Delta s| > 0.11$.}
\label{T:deltas}
\end{center}
\end{table}

\subsection{Change in $H(z)$}
So far the only tests of dark energy proposed have used either
the growth factor or the redshift luminosity distance. If one wishes
to extract $w(z)$, then both of these involve a double integral over
this quantity and degeneracies arise. A more direct and still in
principle observable way is to measure the Hubble parameter
$H(z)$, which is related to $w(z)$ through a single integral.
One way to measure it is to have a characteristic feature
fixed in comoving coordinates which is observed in redshift space.
The relation between redshift space and comoving space is determined
by $H(z)$ and so observing the feature as a function
of redshift determines its evolution. The problem of course is that
there are no characteristic features imprinted, since the distribution of
structure
in the universe is stochastic in nature.

One must therefore look for a characteristic
scale in correlations between structures.
In principle such features could be
provided by baryonic oscillations imprinted in the matter power spectrum, but
in practice this is a weak effect limited to very large scales and
so cannot be made very precise. One is thus left with the
variations in the correlation function slope as a function of scale.
The slope varies from $n\sim 1$
on large scales to $n \sim -3$; on Ly-$\alpha$ forest scales we 
find $\alpha_{\rm eff}=-0.22$  
in CDM models. Hence the scale at which the
slope takes a specific value can be viewed as a standard ruler and
can be traced with redshift. If this slope is measured in redshift space,
then one is measuring
directly $H(z)$. To be able to do this one has to detect the curvature
in the slope over the dynamic range of observations. This is
challenging, since the dynamic range is narrow and the error on the
slope will be large. We find that such a detection should be
possible with the current sample, but is not expected to improve the
constraints on $w(z)$ significantly.

\section{Discussion and Conclusions}\label{S:discussion}

Our results show that the statistical precision one can achieve 
from Ly-$\alpha$ forest using SDSS data is truly impressive. 
The overall amplitude of the power spectrum, its slope and its 
curvature can all be determined with 1-3\% precision. 
Our results are significantly more optimistic than those in
\cite{2001astro.ph..11230Z}, which focused on a higher resolution but smaller 
data sample available from Keck. 
This is primarily a consequence 
of the large sample size of SDSS data, which increases the statistical 
precision on larger scales and moves the pivot point from $0.03$ to 
$0.01$ s km$^{-1}$. 
On larger scales the sensitivity of the power 
spectrum to the gas temperature-density relation is reduced so one is 
measuring more directly the underlying matter power spectrum. 
In addition, 
we have shown here that higher order correlations break the 
degeneracy between the mean flux and the amplitude (this 
degeneracy is broken to some extent already by the power spectrum 
alone). 
Using the power spectrum alone and the bispectrum alone are two
independent ways to confirm this determination.  
Using them together gives even lower error bounds due to the breaking of degeneracies.
The statistical precision of such a data set is competitive with 
the one from the CMB using the MAP and Planck satellites. More 
importantly, combining the two probes will allow one to 
determine the primordial power spectrum to a high 
precision over 3 decades in scale ($0.001<k<2$ $h$Mpc$^{-1}$). 

Furthermore, since one is measuring the power spectrum over a range 
of redshifts ($2<z<4$), one can also study the growth factor 
evolution. While for standard cosmological constant 
models the universe is effectively Einstein-de Sitter for $z>2$, 
dynamical models with rolling scalar fields often 
produce an equation of state 
increasing with redshift (this is especially true for many of the so-called
tracker models). In this case the dark energy or quintessence is 
still dynamically important for $z>2$ and can affect the growth 
rate of perturbations and the Hubble parameter $H(z)$. 
While there is no
simple single parameter combination that describes the sensitivity
to dark energy, it is clear that the precision
is correlated
with $\Omega_q$ at $z>2$. Our results show that if
$\Omega_q(z=2.6)>0.2$ then the deviations in the growth
factor are sufficiently large to be detected in Ly-$\alpha$ forest
spectra using the current SDSS sample. With the full SDSS sample
this limit can be improved further and models with $\Omega_q(z=2.6)>0.1$
should be detectable.
This sensitivity can increase further 
if one can extend the available data set to even lower redshifts by 
using space based observations.

While it is clear that the statistical power of upcoming data sets 
is impressive, the main remaining issue is whether the simplest 
model of the QSO absorption adopted here is valid, or whether there are other 
more complicated astrophysical processes that can 
spoil the picture. Since the expected statistical precision 
is so high one must investigate processes that affect 
the statistics even at a very small (1\%) level. 
There are several possible 
complications that may have an effect and we mention some here: 
metal line absorption along the line of sight (e.g. C-IV, Si-III...), 
an inhomogeneous UV background, kinematic gas outflows generated 
by supernovae from the galaxies, temperature fluctuations, etc. 

Fortunately, while there are several possible contaminations, there
are also several possible tests one can apply to identify and 
remove the contamination. 
Some of these processes can be investigated with the statistics 
used in this paper, such as the power spectrum and bispectrum. In 
the simplest model the main driver of correlations is gravity, 
which induces a very specific relation between low and higher 
order statistics. Any astrophysical contamination is likely to 
destroy these relations. 
Other tests such as the probability distribution of the flux (as a function 
of scale) will also yield useful information \citep{2000ApJ...543....1M}. Another way to 
investigate these effects is through the cross-correlation between 
the forest and galaxies at the same redshift \citep{2003ApJ...584...45A};
%%%This part added because of referee's comment
for theoretical attempts to interpret this result, see \citet{2002ApJ...580...42M}, \citet{2002ApJ...580..634C},  \citet{2002astro.ph.12355K} 
and \citet{2002astro.ph.12126B}. 

If the simplest model passes this and 
other high precision tests it will receive an 
important confirmation that will strengthen the credibility of the results. 
In the opposite case one can still identify the contaminant and apply 
the corrections assuming the contamination is sufficiently well 
understood. This parallels the situation in the CMB, where secondary 
processes and foregrounds are searched for and, if identified, 
subtracted from the primary CMB. It is clear however that a 
validation of the interpretation set forth by the simplest model 
will require a lot of coordinated effort from several groups and 
more effort should be put into this. Our results suggest that 
this is well worth the effort and that the Ly-$\alpha$ forest could 
be our next cosmological gold-mine.

R.M. is supported by an NSF Graduate Research Fellowship.  
This work was supported by NASA, NSF CAREER, David and Lucille Packard 
Foundation and Alfred P. Sloan Foundation.  We thank Nick Gnedin for
the HPM code.  The simulations were performed at the National Center
for Supercomputing Applications.

\bibliography{apjmnemonic,cosmo,cosmo_preprints}

\begin{thebibliography}{}

\bibitem[\protect\citeauthoryear{{Adelberger} et~al.}{{Adelberger}
  et~al.}{2003}]{2003ApJ...584...45A}
{Adelberger} K.~L., {Steidel} C.~C., {Shapley} A.~E.,  {Pettini} M., 2003,
  \apj, 584, 45

\bibitem[\protect\citeauthoryear{{Bruscoli} et~al.}{{Bruscoli}
  et~al.}{2002}]{2002astro.ph.12126B}
{Bruscoli} M., {Ferrara} A., {Marri} S., {Schneider} R., {Maselli} A.,
  {Rollinde} E.,  {Aracil} B., 2002

\bibitem[\protect\citeauthoryear{{Cen} \& {McDonald}}{{Cen} \&
  {McDonald}}{2002}]{2002ApJ...570..457C}
{Cen} R.,  {McDonald} P., 2002, \apj, 570, 457

\bibitem[\protect\citeauthoryear{{Cen} et~al.}{{Cen}
  et~al.}{1994}]{1994ApJ...437L...9C}
{Cen} R., {Miralda-Escude} J., {Ostriker} J.~P.,  {Rauch} M., 1994, \apjl, 437,
  L9

\bibitem[\protect\citeauthoryear{{Cen} et~al.}{{Cen}
  et~al.}{2002}]{2002astro.ph..3524C}
{Cen} R., {Ostriker} J.~P., {Prochaska} J.~X.,  {Wolfe} A.~M., 2002, ApJ,
  submitted (astro-ph/0203524)

\bibitem[\protect\citeauthoryear{{Cen} et~al.}{{Cen}
  et~al.}{2001}]{2001ApJ...559L...5C}
{Cen} R., {Tripp} T.~M., {Ostriker} J.~P.,  {Jenkins} E.~B., 2001, \apjl, 559,
  L5

\bibitem[\protect\citeauthoryear{{Croft} et~al.}{{Croft}
  et~al.}{2002a}]{2002ApJ...580..634C}
{Croft} R.~A.~C., {Hernquist} L., {Springel} V., {Westover} M.,  {White} M.,
  2002a, \apj, 580, 634

\bibitem[\protect\citeauthoryear{{Croft} et~al.}{{Croft}
  et~al.}{2002b}]{2002ApJ...581...20C}
{Croft} R.~A.~C., {Weinberg} D.~H., {Bolte} M., {Burles} S., {Hernquist} L.,
  {Katz} N., {Kirkman} D.,  {Tytler} D., 2002b, \apj, 581, 20

\bibitem[\protect\citeauthoryear{{Croft} et~al.}{{Croft}
  et~al.}{1998}]{1998ApJ...495...44C}
{Croft} R.~A.~C., {Weinberg} D.~H., {Katz} N.,  {Hernquist} L., 1998, \apj,
  495, 44

\bibitem[\protect\citeauthoryear{{Croft} et~al.}{{Croft}
  et~al.}{1999}]{1999ApJ...520....1C}
{Croft} R.~A.~C., {Weinberg} D.~H., {Pettini} M., {Hernquist} L.,  {Katz} N.,
  1999, \apj, 520, 1

\bibitem[\protect\citeauthoryear{{Gnedin} \& {Hui}}{{Gnedin} \&
  {Hui}}{1998}]{1998MNRAS.296...44G}
{Gnedin} N.~Y.,  {Hui} L., 1998, \mnras, 296, 44

\bibitem[\protect\citeauthoryear{{Hamilton}}{{Hamilton}}{2001}]{2001MNRAS.322.%
.419H}
{Hamilton} A.~J.~S., 2001, \mnras, 322, 419

\bibitem[\protect\citeauthoryear{{Hernquist} et~al.}{{Hernquist}
  et~al.}{1996}]{1996ApJ...457L..51H}
{Hernquist} L., {Katz} N., {Weinberg} D.~H.,  {Jordi} M., 1996, \apjl, 457, L51

\bibitem[\protect\citeauthoryear{{Hui} et~al.}{{Hui}
  et~al.}{2001}]{2001ApJ...552...15H}
{Hui} L., {Burles} S., {Seljak} U., {Rutledge} R.~E., {Magnier} E.,  {Tytler}
  D., 2001, \apj, 552, 15

\bibitem[\protect\citeauthoryear{{Kollmeier} et~al.}{{Kollmeier}
  et~al.}{2002}]{2002astro.ph.12355K}
{Kollmeier} J.~A., {Weinberg} D.~H., {Dave'} R.,  {Katz} N., 2002, ArXiv
  Astrophysics e-prints, 12355

\bibitem[\protect\citeauthoryear{{McDonald}}{{McDonald}}{2001}]{2001astro.ph..%
8064M}
{McDonald} P., 2001, ApJ, in press (astro-ph/0108064)

\bibitem[\protect\citeauthoryear{{McDonald}, {Miralda-Escud{\' e}}, \&
  {Cen}}{{McDonald} et~al.}{2002}]{2002ApJ...580...42M}
{McDonald} P., {Miralda-Escud{\' e}} J.,  {Cen} R., 2002, \apj, 580, 42

\bibitem[\protect\citeauthoryear{{McDonald} et~al.}{{McDonald}
  et~al.}{2001}]{2001ApJ...562...52M}
{McDonald} P., {Miralda-Escud{\' e}} J., {Rauch} M., {Sargent} W.~L.~W.,
  {Barlow} T.~A.,  {Cen} R., 2001, \apj, 562, 52

\bibitem[\protect\citeauthoryear{{McDonald} et~al.}{{McDonald}
  et~al.}{2000}]{2000ApJ...543....1M}
{McDonald} P., {Miralda-Escud{\' e}} J., {Rauch} M., {Sargent} W.~L.~W.,
  {Barlow} T.~A., {Cen} R.,  {Ostriker} J.~P., 2000, \apj, 543, 1

\bibitem[\protect\citeauthoryear{{Seljak} \& {Zaldarriaga}}{{Seljak} \&
  {Zaldarriaga}}{1996}]{1996ApJ...469..437S}
{Seljak} U.,  {Zaldarriaga} M., 1996, \apj, 469, 437

\bibitem[\protect\citeauthoryear{{Tegmark}, {Taylor}, \& {Heavens}}{{Tegmark}
  et~al.}{1997}]{1997ApJ...480...22T}
{Tegmark} M., {Taylor} A.~N.,  {Heavens} A.~F., 1997, \apj, 480, 22

\bibitem[\protect\citeauthoryear{{Theuns} et~al.}{{Theuns}
  et~al.}{1998}]{1998MNRAS.301..478T}
{Theuns} T., {Leonard} A., {Efstathiou} G., {Pearce} F.~R.,  {Thomas} P.~A.,
  1998, \mnras, 301, 478

\bibitem[\protect\citeauthoryear{{Vogt} et~al.}{{Vogt}
  et~al.}{1994}]{1994SPIE.2198..362V}
{Vogt} S.~S. et~al., 1994, in Proc. SPIE Instrumentation in Astronomy VIII,
  David L. Crawford; Eric R. Craine; Eds., Volume 2198, p. 362, Vol. 2198

\bibitem[\protect\citeauthoryear{{Zaldarriaga}, {Scoccimarro}, \&
  {Hui}}{{Zaldarriaga} et~al.}{2001a}]{2001astro.ph..11230Z}
{Zaldarriaga} M., {Scoccimarro} R.,  {Hui} L., 2001, ApJ, submitted
  (astro-ph/0111230)

\bibitem[\protect\citeauthoryear{{Zaldarriaga}, {Seljak}, \&
  {Hui}}{{Zaldarriaga} et~al.}{2001b}]{2001ApJ...551...48Z}
{Zaldarriaga} M., {Seljak} U.,  {Hui} L., 2001, \apj, 551, 48

\bibitem[\protect\citeauthoryear{{Zhang}, {Anninos}, \& {Norman}}{{Zhang}
  et~al.}{1995}]{1995ApJ...453L..57Z}
{Zhang} Y., {Anninos} P.,  {Norman} M.~L., 1995, \apjl, 453, L57

\end{thebibliography}
   \bibliographystyle{mnras}

\end{document}